\documentclass[10pt]{SolarPhysics}

\usepackage{graphicx}
\usepackage{psfig}

\newcommand{\be}{\begin{equation}}
\newcommand{\ee}{\end{equation}}
\newcommand\eq{Equation}
\newcommand\eqs{Equations}
\newcommand\fig{Figure}
\newcommand\figs{Figures}

\newcommand\etal{{\it et al.}}
\newcommand\ie{{\it i.e.}}

\newcommand{\half}{\hbox{${1\over2}$}}

\newcommand{\zhat}{{\bf \hat{z}}}

\newcommand{\bvec}{{\bf B}}

\newcommand{\bc}{{\cal C}}
\def\res{\mathop{{\rm Res}}}
\newcommand\Bxfr{B^{\rm (fr)}_x}
\newcommand\Byfr{B^{\rm (fr)}_y}
\newcommand\Bxs{B_{0,x}}
\newcommand\Bys{B_{0,y}}

\begin{document}

\begin{article}
\begin{opening}
\title{Breakout and Tether-cutting Eruption Models Are Both Catastrophic (Sometimes)}

\author{D.W. \surname{Longcope}$^1$ and T.G. \surname{Forbes}$^2$}
\institute{$^1$ Department of Physics, Montana State University, \\
  Bozeman, MT 59717\\
  $^2$ Institute for the Study of Earth, Oceans, and Space (EOS)\\
  University of New Hampshire, 
  Durham, NH 03824}


\begin{abstract}
We present a simplified analytic model of a quadrupolar magnetic field and flux rope to model coronal mass ejections.  The model magnetic field is two-dimensional, force-free and has current only on the axis of the flux rope and within two currents sheets.  It is a generalization of previous models containing a single current sheet anchored to a bipolar flux distribution.  Our new model can undergo quasi-static evolution due either to changes at the boundary or to magnetic reconnection at either current sheet.  We find that all three kinds of evolution can lead to a catastrophe known as loss of equilibrium.  Some equilibria can be driven to catastrophic instability either through reconnection at the lower current sheet, known as {\em tether cutting}, or through reconnection at the upper current sheet, known as 
{\em breakout}.  Other equilibria can be destabilized through only one and not the other.  Still others undergo no instability, but evolve increasingly rapidly in response to slow steady driving (ideal or reconnective).  One key feature of every case is a response to reconnection different from that found in simpler systems.  In our two-current sheet model a reconnection electric field in one current sheet causes the current in that sheet to {\em increase} rather than decrease.  This suggests the possibility for the microscopic reconnection mechanism to run away.
\end{abstract}

\date{Draft: \today}

\end{opening}

\section{Introduction}

Coronal mass ejections (CMEs) are believed to occur through the sudden conversion of magnetic energy into bulk kinetic energy as well as heating and particle acceleration 
\cite{Forbes2000,Klimchuk2001,Lin2003,Moore2006}.  Models of these events generally include a twisted flux rope, or sheared magnetic arcade, above a distribution of photospheric magnetic flux.  Current carried by the flux rope or sheared arcade is the source of free magnetic energy, and eruption occurs as this energy is released through an upward expansion, and diminishment, of the current.  

This basic CME model has been studied through simplified analytic calculation 
\cite{Low1977,VanTend1978,Martens1989,Forbes1991,Isenberg1993,Forbes1995,Titov1999,Isenberg2007} 
as well as sophisticated numerical simulation 
\cite{Mikic1994b,Antiochos1999,Amari2000,Manchester2007,Roussev2012,Karpen2012}.  In spite of this substantial effort there remains uncertainty as to what role magnetic reconnection might play in an eruption.  In some models, termed {\em ideal} by Forbes (2000),\nocite{Forbes2000} the eruption is initiated by an ideal, current driven instability, but reconnection can occur as a consequence.  In the alternative class of models, termed {\em resistive}, the cause of the eruption is magnetic reconnection occurring at some current sheet which would be stable in the absence of resistivity.  Reconnection is present in both kinds of model, however, in one it is the driver and in the other it is a secondary effect.  It is notoriously difficult to distinguish between cause and effect in an experiment or a numerical simulation.  Such questions are therefore better answered through simplified analytic treatments.

\begin{figure}[htp]
\psfig{file=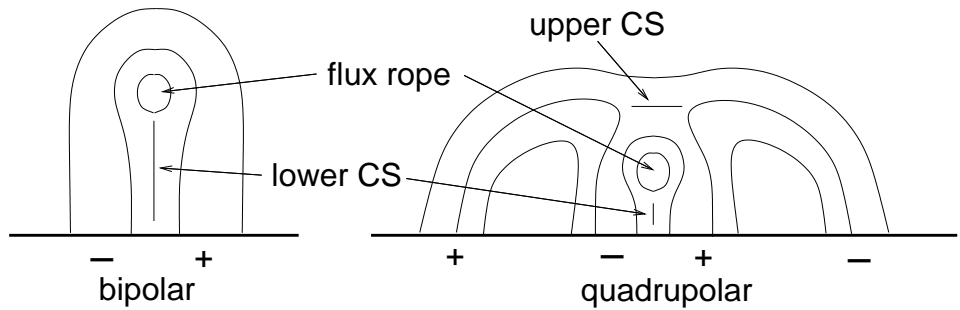,width=5.0in}
\caption{Generic CME models after Moore and Sterling (2006).  In the {\em bipolar} model (left) the photospheric flux distribution consists of one positive and two negative regions, denoted by $+$s and $-$s respectively.  Field lines connecting these are shown as thin solid curves.  The {\em quadruplar} model (right) includes two positive and two negative sources.  Current sheets form below the flux rope in both models, shown as darker straight lines.  The quadrupolar model includes a second current sheet above the flux rope.}
	\label{fig:Moore_toon}
\end{figure}

Forbes and Isenberg (1991)\nocite{Forbes1991} proposed an analytic model of a CME capable of addressing these questions.  The model assumed a two-dimensional magnetic field including a flux rope idealized as a line current.  The field was force-free and current-free with the exception of the line current and a current sheet directly beneath it, as illustrated on the left of\ \fig\ \ref{fig:Moore_toon}.  It is possible to characterize the complete equilibrium space of such magnetic configurations.  Quasi-static evolution occurs as changing boundary conditions or magnetic reconnection causes the system to move through the equilibrium space.  Forbes and Isenberg (1991)\nocite{Forbes1991} found some evolutionary scenarios in which the system encountered a catastrophe and no nearby equilibrium existed to evolve to.  This situation represents a kind of instability termed a {\em loss of equilibrium} (LOE).  Under dynamical evolution the flux rope will move rapidly upward toward a new equilibrium position.

The model of Forbes and Isenberg (1991) was aimed only at elucidating the nature of the instability ultimately responsible for eruption.  It is therefore not really a model of the eruption, but is a model of the {\em pre-eruption} equilibrium.
To this end it neglects many important aspects of real CMEs such as their dynamics and acceleration, and the conversion of magnetic energy.  The model of the pre-eruption equilibrium assumes magnetic energy to be dominant and therefore neglects all other contributions, thermal, kinetic and gravitational potential energy.  In order to maintain analytic tractability, the magnetic field is assumed take a simplified form, including a rigid conducting cylinder (the line current) to represent the current-carrying flux rope.  This model element, used in place of a tube of axial flux and axial current, is a passive element in the magnetic equilibrium, doing no work on the surrounding field, but it does add enough complexity to permit instability.  In spite of this simplification, their model has proven useful as a building block for many subsequent studies, including an investigation of the light curves of long-duration flares \cite{Reeves2005}.

When it undergoes LOE in the model, the flux rope makes a sudden upward motion, but does not completely erupt from the Sun.  There is always a stable equilibrium with the flux rope at finite height.  The state of full eruption (the flux rope moved to infinity) has infinite magnetic energy as long as some flux connects over the flux rope, as in \fig\ \ref{fig:Moore_toon}.  This is a manifestation of the {\em Aly-Sturrock conjecture} \cite{Sturrock1991,Aly1991}, which states that a field with closed field lines will always have lower magnetic energy than a field with open field lines anchored to the same photospheric flux distribution.  It is therefore necessary for magnetic reconnection to eliminate all overlying flux in order to achieve full eruption in the analytic model of Forbes and Isenberg (1991).\nocite{Forbes1991}

The analytic model points to two possible roles that reconnection might play in driving an eruption.  It can {\em trigger} the eruption by bringing the system to the point of catastrophe where mechanical equilibrium is lost.  Once the eruption is triggered reconnection plays no further role and the system evolves ideally on Alfv\'enic time scales to its new equilibrium.  Alternatively, the reconnection can actually drive the evolution as it must to achieve full eruption {\em after} LOE.  In this case the rate of evolution scales with the reconnection electric field in some manner.  

Motivated by observations, and as a means of circumventing the Aly-Sturrock conjecture,\footnote{Rather than opening the field lines overlying the flux rope it is possible to remove by closing them across the PILs on either side \cite{Antiochos1999}.} more recent modeling has focussed on photopsheric flux distributions more complex than the bipole of 
Forbes and Isenberg (1991).\nocite{Forbes1991}  Several models have considered a quadrupolar flux distribution in which the photospheric field has three separate polarity inversion lines (PILs; Antiochos 1998; Moore \etal\ 2001)\nocite{Antiochos1998,Moore2001}.  In the presence of this more complex photospheric distribution there will be current sheets both above and below the flux rope, as shown in \fig\ \ref{fig:Moore_toon}.  Reconnection can occur at either or both current sheets, and can be either a cause or an effect of the eruption.  Moore and Sterling (2006)\nocite{Moore2006} identify three possible scenarios, two resistive and one ideal, by which the eruption might occur.  Resistive eruption can be triggered either by reconnection at the lower current sheet ({\em internal tether cutting}) or by reconnection at the upper current sheet ({\em breakout}).  A recent large-scale numerical simulation by Karpen, Antiochos, and DeVore (2012)\nocite{Karpen2012} shows reconnection occurring at both sites during the eruption.  A set of auxiliary simulations permit the conclusion that the breakout reconnection is the first to occur, and is responsible for subsequent reconnection at the lower current sheet, they call {\em flare reconnection}.  These experiments also suggest that the eruption is either triggered or driven by that reconnection: it is a resistive rather than an ideal instability.

A deeper understanding of the role played by reconnection in this more complex geometry could be achieved through a generalization of the analytic model of Forbes and Isenberg (1991).\nocite{Forbes1991}  While Isenberg, Forbes, and 
D\'emoulin (1993) and Forbes, Priest, and Isenberg (1994)\nocite{Forbes1994} do consider more complex flux distributions, including a quadruplar case, they assumed only a single current sheet and therefore could not distinguish between breakout or tether-cutting reconnection.   A semi-analytic study by Zhang, Hu, and Wang (2005)\nocite{Zhang2005} of a quadrupolar equilibrium with two current sheets in a spherical geometry found a complex ``double catastrophe'' under a single evolutionary sequence.  Unfortunately their method, unlike the fully analytic method of Forbes and Isenberg (1991),\nocite{Forbes1991} was not amenable to an exhaustive exploration of parameter space, and the double catastrophe is still not well understood.  \cite[did, however, confirm it using a time-dependent numerical simulation.]{Zhang2006b}

In the present work we develop an analytic, two-dimensional model of the configuration on the right of \fig\ \ref{fig:Moore_toon}, with a quadrupolar photospheric distribution, a single flux rope, and two current sheets.  We find possibilities for LOE in direct analogy to Forbes and Isenberg (1991).\nocite{Forbes1991}  This can occur in any of the three ways enumerated by Moore and Sterling (2006):\nocite{Moore2006} reconnection at either the upper or lower current sheets or through ideal motion alone.  We find some configurations where LOE can occur through one kind of reconnection but {\em not} the other.  We also find situations in which no LOE occurs, yet the flux rope is driven upward extremely rapidly.  In spite of the quadrupolar flux distribution, the Aly-Sturrock conjecture still prevents full eruption except through reconnection at the upper current sheet.

\section{The Model Field}

\subsection{The Complex Potential}

Following Forbes and Isenberg (1991)\nocite{Forbes1991} we express the two-dimensional magnetic field using a complex function, $B_y + i B_x ~=~ \hat{F}(x+iy)$, of the complex spatial coordinate $w=x+iy$.  The use of complex functions to describe certain two-dimensional magnetic equilibria is a well-known technique \cite{Green1965,Syrovatskii1971,Priest1975,Tur1976,Aly1989,Parnell1994}, of which Priest \& Forbes (2000, Section 2.2.1) provide a particularly lucid exposition.  A field generated this way has zero divergence and zero current wherever the function $\hat{F}(w)$ is analytic.  A simple pole corresponds to a line current if it has a real residue or a line charge if the residue is imaginary.  Current sheets are generated by branch cuts provided the field is everywhere tangent to the cut \cite{Aly1989}.

We begin by assuming a photospheric quadrupole consisting of four photospheric ($y=0$) line charges, P1, N2, P2 and N1, located at $x=-x_1,\,-x_2,\,+x_2,\,+x_1$ respectively and with charge (per ignorable length) $+Q_1$, $-Q_2$, $+Q_2$ and $-Q_1$.  We assume that, except at the sources, the magnetic field at the photosphere is purely horizontal, so $\hat{F}$ is purely imaginary there.  In addition a line current carrying net current $I_0$ is located at $(x,y)=(0,h)$, to represent a flux rope or {\em current filament}.  In order to prevent this from contributing vertical field at $y=0$ we must add an image current, $-I_0$, at $(x,y)=(0,-h)$.  The coronal field from this arrangement which is current-free except for the line current, is generated by complex function
\begin{eqnarray}
  \hat{F}^{(X)}(w) &=& {iQ_1\over w+x_1} - {iQ_2\over w+x_2}
  + {iQ_2\over w-x_2} - {i Q_1\over w - x_1}
  + {2I_0\over w - ih} - {2I_0\over w+ih} \nonumber \\
  &=& -{2i Q_1x_1\over w^2 - x_1^2 } ~+~  {2i Q_2x_2\over w^2 - x_2^2 }
  ~+~  {4 iI_0 h\over w^2 + h^2} ~~,
  	\label{eq:Fpot}
\end{eqnarray}
which is clearly a sum of the six contributions described above.  

Equation (\ref{eq:Fpot}) contains a singularity at $w=ih$, the line current, but we exclude this region from our model.  Following Forbes and Isenberg (1991) we assume the current flows on the surface of a rigid cylindrical conductor of radius $R\ll h$.  This conductor represents a flux rope with axial flux and axial current distributed in some manner throughout the cylindrical volume.     Isenberg, Forbes, and D\'emoulin (1993)\nocite{Isenberg1993} investigated a version of the same model with a self-consistent internal distribution of axial and azimuthal flux throughout the flux rope, responsive to forces from the external magnetic field.  
In the end, however, this complication led only to a relation between $R$ and the net current $I_0$, in place of a radius assumed constant.  This ultimately made little difference to the structure of the equilibrium space, including the LOE, provided the radius was small ($R\ll h$).  We use here the case of a rigid conductor both for simplicity and to provide the conceptual advantage of doing no work on the external field, and therefore contributing nothing to the equilibrium energy.  

Due to its small radius the field outside the conductor is well approximated by \eq\ (\ref{eq:Fpot}) in the spirit of a multi-pole expansion \cite{Isenberg1993}.  
Because the radius is non-zero, the field strength, and therefore magnetic energy, is never infinite.

To leading order in $R/h$, the Lorentz force on the current-carrying conductor can be computed from the magnetic field, excluding self-field, which would occur at the cylinder's center 
\be
  \Byfr+i\Bxfr ~=~ {d\over dw}\left[(w-ih)\hat{F}(w)\right]_{w=ih} 
  ~=~  {2i Q_1x_1\over h^2 +x_1^2 } -  {2i Q_2x_2\over h^2+x_2^2 }
  ~-~  {I_0\over ih} ~~;
\ee
(the super-script referring to ``flux rope''.)
This expression is purely imaginary, and thus the field  is purely horizontal, due to the left-right symmetry of the arrangement.  It consists of the contribution of the photospheric sources, $\Bxs$, and of the image current
\be
  \Bxfr ~=~ {2Q_1x_1\over h^2 +x_1^2 } -  {2Q_2x_2\over h^2+x_2^2 } ~+~ {I_0\over h}
  ~=~ \Bxs(h) ~+~ {I_0\over h} ~~.
  	\label{eq:Bxlc}
\ee
For the conductor to be in force balance this field must vanish, meaning the current should be
\be
  I_0(h) ~=~ -h\Bxs(h) ~=~ -h\left[
  {2Q_1x_1 \over h^2+x_1^2} ~-~ {2Q_2x_2 \over h^2+x_2^2} \right] ~~.
  	\label{eq:Ih}
\ee

Field lines can be found from contours of the flux function, $A(x,y)$, defined so that $\bvec=\nabla A\times\zhat$.  When the field is generated from complex function $\hat{F}(w)$, the flux function is the real part of the integral
\be
  \hat{A}(w) ~=~ -\int_0^{w}\hat{F}(w')\, dw ~~,
  	\label{eq:Aw_int}
\ee
performed along any path remaining within the corona, $y>0$.  The real part of this integral will be single-valued (\ie\ path-independent) provided all line charges (poles with imaginary residues) are confined to the periphery of the domain, as they are in our configuration.  Integrating the function from \eq\ (\ref{eq:Fpot}), gives a complex potential
\be
  \hat{A}^{(X)}(w) ~=~ iQ_1\ln\left({x_1-w\over x_1+w}\right) ~-~
  iQ_2\ln\left({x_2-w\over x_2+w}\right) ~+~ 2I_0\ln\left({ih-w\over ih+w}\right) ~~,
  	\label{eq:A0}
\ee
with singularities at each pole and at the line current.

\begin{figure}[htp]
\centerline{\psfig{file=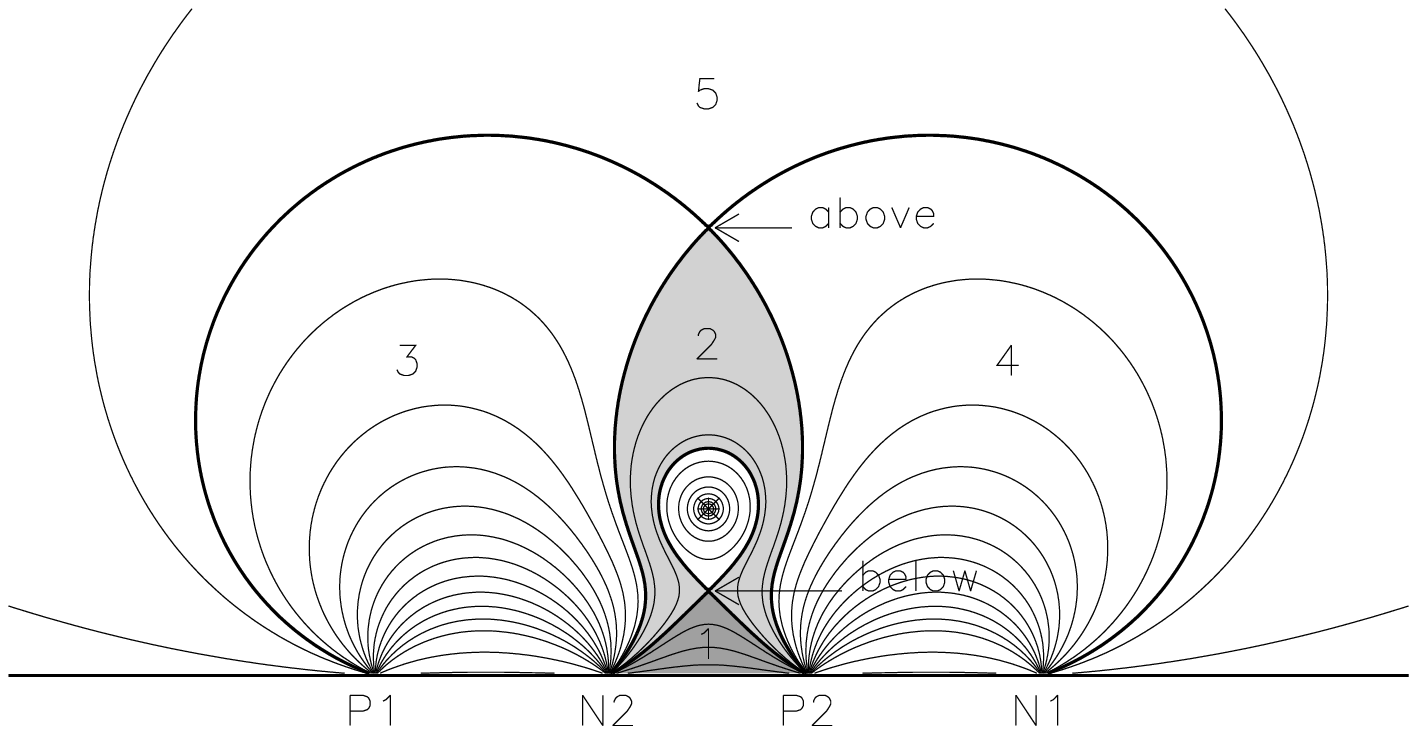,width=4.0in}\psfig{file=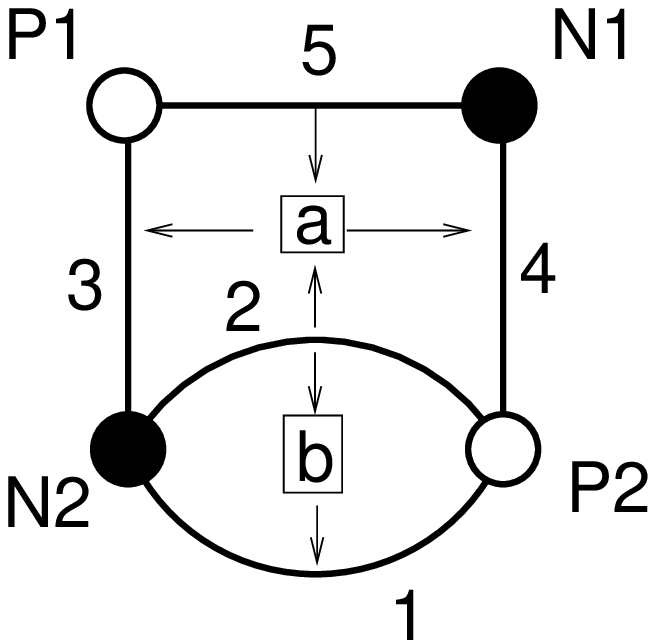,width=1.1in}}
\caption{The field lines for the X-point field given by \eq\ (\ref{eq:A0}).  The flux rope consists of closed contours encircling the line current (asterisk) at $y=h=0.5x_1$.  Outside of these closed field lines all others connect a positive and negative photospheric source.  There are five distinct connectivities forming five labelled regions.  Two different regions connect P2 to N2 and are colored dark grey (region 1) and light grey (region 2).  The connectivities of the regions are depicted schematically as a graph on the right.  Two X-points, labelled ``above'' and ``below'' create the separatrices between the regions.  Their relation to the field topology is indicated by boxed labels in the graph.}
	\label{fig:A0}
\end{figure}

Figure \ref{fig:A0} shows contours of ${\rm Re}(\hat{A}^{(X)})$ for the case $Q_2=1.2Q_1$, $x_2=0.3x_1$, and $h=0.5x_1$.  The function has saddle points at $y=y_a$ and $y=y_b$, above and below the line current.    The complex field given by \eq\ (\ref{eq:Fpot}) can be written
\be
  \hat{F}^{(X)}(w)~=~ iD_0{(w^2 + y_a^2)(w^2+y_b^2)\over (w^2 - x_1^2)(w^2 - x_2^2)
  (w^2 + h^2)} ~~,
  	\label{eq:F0}
\ee
where $D_0= 2(Q_2x_2-Q_1x_1+2I_0h)$ is the overall dipole moment and $y_a^2$ and $y_b^2$ are the roots of the quadratic
\be
  Q_1x_1( y^2 + x_2^2 )( y^2 - h^2 ) - Q_2x_2( y^2 + x_1^2 )( y^2 - h^2 ) 
  - 2I_0 h ( y^2 + x_1^2)( y^2 + x_2^2 ) = 0 ~~.
  	\label{eq:quadratic}
\ee
We call this field the X-point field, since it contains X-points at $w=iy_a$ and $w=iy_b$; the superscript on $\hat{F}$ and $\hat{A}$ refer to this property.

\subsection{Domain Fluxes}

Contours at the saddle values, 
$A_a=A(0,y_a)$ and $A_b=A(0,y_b)$, form separatrices dividing field lines into five connectivity domains (labelled 1--5) and the flux rope consisting of closed field lines.  The graph on the right of \fig\ \ref{fig:A0} shows how field lines from the five domains connect the four sources.  There are two distinct classes of field lines interconnecting sources P2 and N2 occupying regions shaded in different grey scales and denoted 1 and 2 --- they are represented by separate edges connecting the same vertices.

The net flux (per ignorable length) in each of the five domains, denoted $\psi_k$, are  
related to the source fluxes, $\pi Q_j$, as well as the saddle values 
$A_a$ and $A_b$.  For each vertex of the graph in \fig\ \ref{fig:A0} the sum of incident edge fluxes must equal the vertex flux,
\be
  \psi_1+\psi_2+\psi_3 ~=~ \pi Q_2 ~~,~~
  \psi_3+\psi_5 ~=~ \psi_4+\psi_5 ~=~ \pi Q_1 ~~.
  	\label{eq:graph}
\ee
This provides three conditions on the five domain fluxes, leaving two degrees of freedom.  This is consistent with the existence of two {\em circuits} in the graph, labeled a and b in \fig\ \ref{fig:A0}.  Each circuit is associated with a separator which in two-dimensional fields are X-points \cite{Longcope2001b}.  Separator a is the ``above'' X-point and 
b is the ``below'' X-point.  The below X-point lies above only domain 1 so $A_b=-\psi_1$, while the above X-point  overlies both domains and $A_a=-(\psi_1+\psi_2)$.

Using the above relations it is possible to deduce all domain fluxes given the values of $A_a$ and $A_b$.  These two values relate directly to the two domain fluxes, $\psi_1=-A_b$ and $\psi_5 = -A_a + \pi(Q_1-Q_2)$, and the other three follow from these, $\psi_3=\psi_4=\pi Q_1-\psi_5$ and $\psi_2 = \pi(Q_2-Q_1)+\psi_5 - \psi_1$.

\begin{figure}[htp]
\centerline{\psfig{file=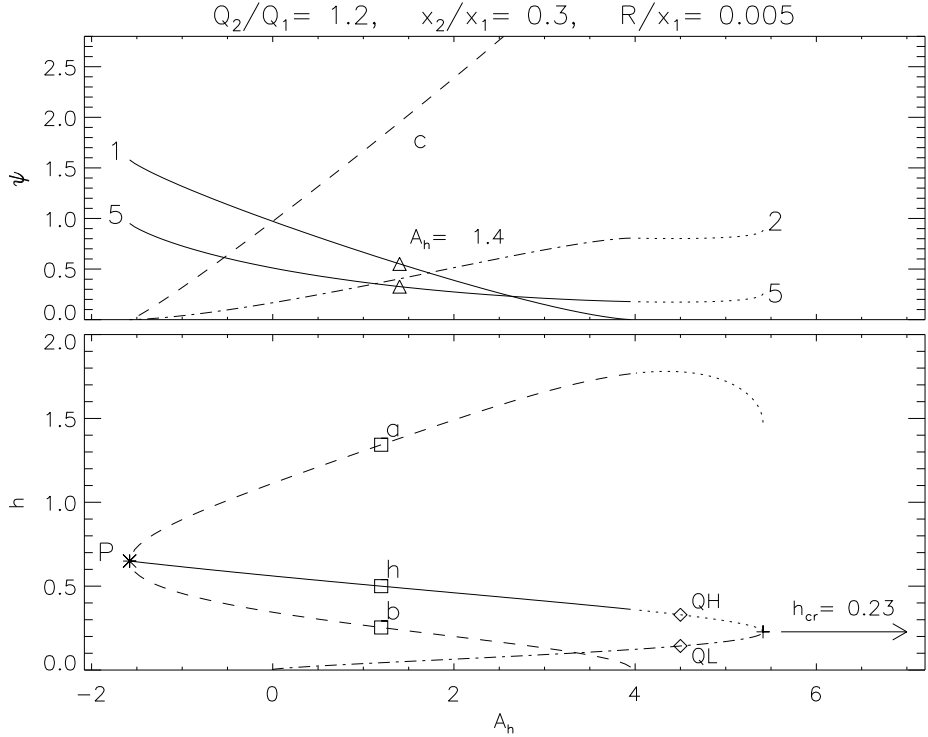,width=5.5in}}
\caption{X-point equilibria for specified values of $A_h$ (normalized to $Q_1$) for current on conductor of radius $R=0.005x_1$.  The bottom panel shows the equilibrium height, $h$, of the current (solid), and the heights of the X-points $y_a$ and $y_b$ (dashed), all normalized to $x_1$.  Squares correspond to the configuration shown in \fig\ \ref{fig:A0}.
Dotted portions of $h$ and $y_a$ curves correspond to the unstable turtle configuration; the broken curve is 
for $h$ in the stable turtle configuration.  Diamonds labeled QL and QH are the stable and unstable (respectively) turtle configurations shown  in \fig\ \ref{fig:turtle}.  The asterisk P is the potential field with $I_0=0$.
The top panel shows fluxes $\psi_1$ and $\psi_5$ (solid), $\psi_2$ (broken) and the closed flux $\psi_c$ (dashed) in the X-point field.}
	\label{fig:Ah}
\end{figure}

The net flux between the flux rope and the photosphere is related to the flux function evaluated at the cylindrical surface of the flux rope.  The flux function is constant along the surface of a perfect conductor, as we have assumed the flux rope to be.  This is true of our field, $A^{(X)}(x,y)$, only to leading order in $R/h\ll1$.  We choose to evaluate the enclosed flux at the conductor's lower edge
\be
  A_h ~=~ A(0,h-R) ~=~ 2Q_1\tan^{-1}(h/x_1) -
  2Q_2 \tan^{-1}(h/x_2) + 2I_0(h)\ln(2h/R) ~~.
  	\label{eq:Ah}
\ee
This consists of $\psi_1$, directed leftward, and a flux $\psi_c$, in closed field lines directed rightward:
$A_h=\psi_c-\psi_1$.  By combining \eqs\ (\ref{eq:Ih}) and (\ref{eq:Ah}) it is possible to derive the location $h$ and  current $I_0$ from a specified value of $A_h$, as illustrated in \fig\ \ref{fig:Ah}.

In this model the flux rope is considered a distinct element that does not exchange flux with the external field.  Following this reasoning we take $A_h$, the flux external to the flux rope, to be a constant.  This is equivalent to replacing the flux rope with a perfect conductor as originally proposed by Forbes and Isenberg (1991).  All dynamical evolution of the coronal field must therefore conserve the value of $A_h$.  The net current in the flux rope will therefore change, in response to vertical displacement,
\begin{eqnarray}
  \left.{\partial I_0\over\partial h}\right)_{A_h} &=& -{1\over \ln(2h/R)}\left(
  {Q_1x_1\over h^2+x_1^2} - {Q_2x_2\over h^2+x_2^2} + {I_0\over h} \right) \nonumber \\
  &=&-{1\over \ln(2h/R)}\left(
  \half \Bxs(h)+{I_0\over h} \right)~~,
  	\label{eq:dIdh}
\end{eqnarray}
where $\Bxs$ is the horizontal field due to the photospheric sources alone.  Had a different assumption been made about the flux rope, there would be a slightly different relation between $I_0$ and $h$ \cite{Isenberg1993}.  The assumption made here, of a rigid conductor, prevents the flux rope from contributing work to the system.  In that sense it is a conservative assumption and we will continue to use it.

\subsection{Stability of the Flux Rope}
\label{sec:stabX}

Force balance requires that the vertical force $F_y=I_0\Bxfr$, and thus the horizontal field experienced by the flux rope, vanish according to \eq\ (\ref{eq:Bxlc}).  The stability of the equilibrium depends on how the force $F_y$ varies as the line current is vertically displaced 
\be
  {dF_y\over dh} ~=~ I_0{d\Bxfr\over dh} ~=~
  I_0\left( {\partial \Bxs\over \partial y} - {I_0\over h^2} + {1\over h}{dI_0\over dh} \,\right) ~~.
\ee
If the displacement is subject only to the conservation of $A_h$, and the field is able to remain in potential form from \eq\ (\ref{eq:F0}), then $dI_0/dh$ can be taken from \eq\ (\ref{eq:dIdh}) yielding
\be
  {dF_y\over dh} ~=~ I_0\left( {\partial \Bxs\over \partial y} +{ \Bxs\over h}
  +{\Bxs\over 2h\ln(2h/R)} \right) ~~,
  	\label{eq:dFydh}
\ee
after evaluating the expression at the equilibrium where $I_0=-h\Bxs$.

Stability to vertical displacement requires $dF_y/dh<0$.  The first term in \eq\ (\ref{eq:dFydh}) must be positive in order to assure stability to {\em horizontal}, displacement since
\be
  {dF_x\over dx} ~=~ -I_0{\partial\Bys\over\partial x} ~=~ -I_0{\partial\Bxs\over\partial y} ~~,
\ee
must also be negative (the coronal current generated by the photospheric sources alone
$\partial\Bxs/\partial y - \partial\Bys/\partial x$ must vanish).  The other two terms in \eq\ (\ref{eq:dFydh}) 
will be stabilizing since $I_0\Bxs=-hI_0^2<0$.  

\begin{figure}[htp]
\centerline{\psfig{file=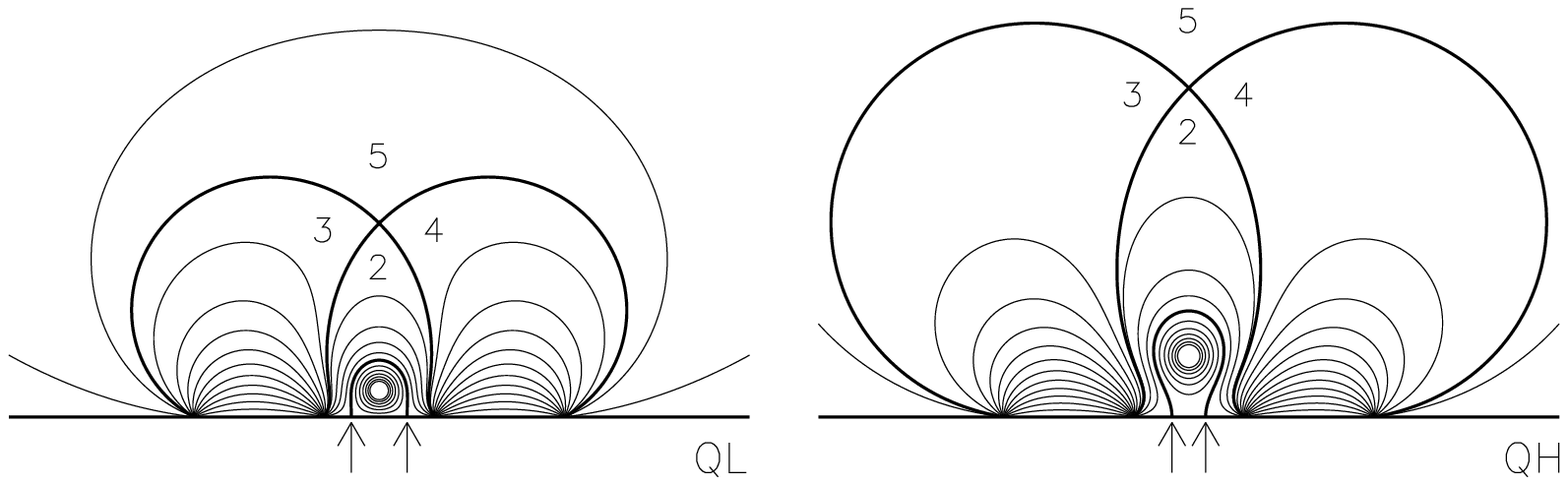,width=4.0in}~~\psfig{file=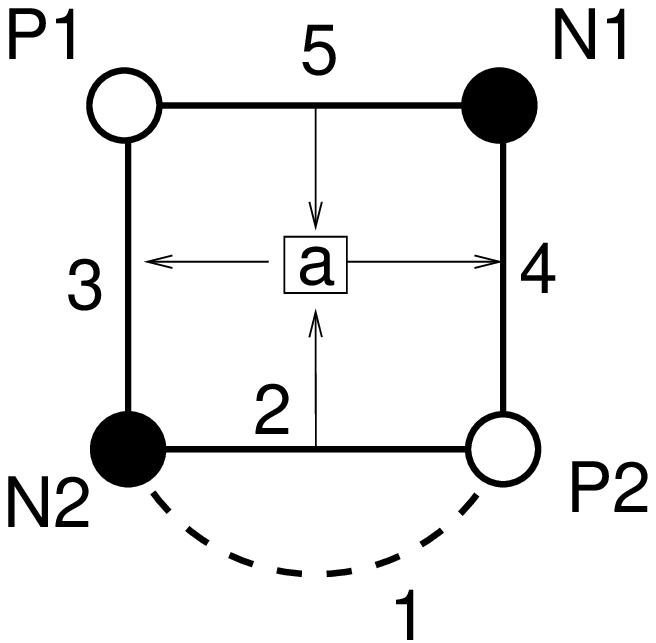,width=1.1in}}
\caption{The X-point fields QL (left) and QH (center) corresponding to the points in \fig\ \ref{fig:Ah} with 
$A_h=4.5Q_1$.  The separatrix overlying the flux rope --- the ``turtle shell'' 
(dark solid curve) is connected to photospheric null points at 
$x=\pm|y_a|$ indicated by arrows.  Each configuration has the same graph (right panel) of four extant domains; domain $\psi_1$ (shown dashed) is missing.}
	\label{fig:turtle}
\end{figure}

In the limit of an ideal line current, $R\to0$, stability requires $\partial[h\Bxs(0,h)]/\partial h<0$ \cite{Forbes1991}.  For the configuration shown in \fig\ \ref{fig:A0}, this would occur at $h=0.21x_1$.  
A finite radius contributes the third term to \eq\ (\ref{eq:dFydh}) which is stabilizing (provided $R<2h$).  This 
raises the critical height in \fig\ \ref{fig:Ah} to the value shown, $h_{\rm cr}=0.23 x_1$.  All the stable configurations, \ie\ where $h<h_{\rm cr}$, such as QL in \fig\ \ref{fig:turtle}, are characterized by a negative root ($y_b^2<0$) of \eq\ (\ref{eq:quadratic}).  In these cases, which  we refer to these as {\em turtle} configurations, 
the lower coronal X-point is replaced by two photospheric X-points at $x=\pm|y_b|$.  
The domain of closed flux contacts the photosphere between these nulls, and so domain $1$ is entirely missing.  This means $\psi_1=0$ and the domain graph is the simpler one shown on the right panel of \fig\ \ref{fig:turtle}.

The right portion of the $h(A_h)$ curve in \fig\ \ref{fig:Ah} is double valued: the upper portion is solid and the lower portion is broken (dash-dot).  This is similar to the catastrophe scenarios considered by Forbes and Isenberg (1991),\nocite{Forbes1991} where the lower branch is stable ($h<h_{\rm cr}$) and the upper branch is unstable ($h>h_{\rm cr}$).  As $A_h$ is increased the two equilibrium heights approach and ultimately merge at $A_h=A_{h,{\rm cr}}=5.4Q_1$ where both values converge at $h=h_{\rm cr}$.  There are no solutions for $A_h>A_{h,{\rm cr}}$, leading to the term {\em loss-of-equilibrium}.

For the case considered here ($R=0.005x_1$) all stable equilibria are of the turtle variety, lacking the fifth domain of connectivity.  Increasing the flux rope radius, $R$, will increase the contribution of the third term in \eq\ (\ref{eq:dFydh}) until, at $R=0.49x_1$, $h_{\rm cr}=0.36x_1$ coincides with the marginal state $y_b^2=0$.  Radii larger than this would provide examples of stable equilibria of the generic quadrupolar variety like that in \fig\ \ref{fig:A0}.  They are, however, unphysical since $R>h$.  We therefore conclude that the photopsheric source distribution $Q_2/Q_1=1.2$ and $x_2/x_1=0.3$ has no stable X-point equilibria of the generic quadrupolar variety.

The marginal cases ($y_b^2=0$) of all possible source distributions (\ie\ all $x_2/x_1$ and $Q_2/Q_1$) has a minimum value $R/h_{\rm cr}=2e^{-1}$, which is not, at least, unphysical (\ie\ $R<h$).  The minimum value occurs for limiting case of a simple dipole ($x_1/x_2\to\infty$) --- the case treated by Forbes, Priest, and Isenberg (1994)\nocite{Forbes1994} who report the same value of $R/h$ for marginal stability.  While physically admissible this ratio is well outside our assumed limit $R\ll h$.  We therefore conclude that no generically quadrupolar X-point fields (\ie\ fields with $\psi_1>0$) 
are stable against vertical displacement of its current filament.

The instability of the equilibrium can be alternatively related to magnetic energy above that of the potential field $\bvec_0=\nabla A_0\times\zhat$, called free magnetic energy
\be
  \Delta W ~=~ {1\over 8\pi}\int\nabla (A-A_0)\cdot\nabla(A+A_0)\, dx\,dy 
  ~=~ I_0^2\ln(2h/R) ~~.
\ee
The change in free magnetic energy with height of the flux rope
\be
  {d(\Delta W)\over dh} ~=~ I_0\left( 2\ln(2h/R){dI_0\over dh} + {I_0\over h}\right) ~=~
  -I_0 \left( \Bxs + {I_0\over h} \right) ~=~ -F_y ~~,
\ee
after replacing $dI_0/dh$ using \eq\ (\ref{eq:dIdh}).  This confirms that all work done moving the flux rope is stored in the magnetic field.  This occurs because there is no electromotive force  in the line current ($-dA_h/dt=0$) and the rigidity of the conductor ($R$ held fixed) means no work is done changing the volume of the flux rope.  Stability would follow from $\Delta W$ being a minimum at the equilibrium $I_0=-h\Bxs$.  As we have shown above, this does not happen for any generically quadrupolar configuration.

This tells us that when the magnetic field is subject to only the single constraint on $A_h$, its energy can be reduced by moving the flux rope vertically from its equilibrium position.  
In order to reduce the energy, and maintain the field in configuration generated by $\hat{F}^{(X)}$, it is necessary to transfer flux across the X-points at $y_a$ and $y_b$.  This transfer would not, however, be permitted by a highly conducting plasma such as the corona.  Instead, each of the X-points would deform into a current sheet in order to preserve the fluxes $\psi_1$ and $\psi_5$ in addition to the total flux $A_h=\psi_c-\psi_1$.  These additional constraints will necessarily {\em increase} the free magnetic energy of the field, and can stabilize the equilibrium.

\section{Equilibria with Current Sheets}

\subsection{Equilibrium Equations}

A non-potential field anchored to the same photospheric sources and containing the same flux rope can be constructed as a {\em flux constrained equilibrium} (FCE; Longcope, 2001)\nocite{Longcope2001b}.  The magnetic energy is minimized subject to boundary conditions and constraints on all domain fluxes.  Following the discussion above we see that provided the field has the same connectivity, \ie\ the graph from \fig\ \ref{fig:A0}, the complete set of domain flux constraints is equivalent to constraints on the flux function's value at the separatrices $A_a$ and $A_b$.  The minimizing field is potential everywhere except at separator current sheets \cite{Longcope2001b}.  

Such a field can be represented by a complex potential $\hat{F}(w)$ containing branch cuts in place of the simple nulls.  A null point $(w-w_0)$ is replaced by the expression, 
$\sqrt{(w-\tau_p)(w-\tau_q)}$ where $\tau_p$ and $\tau_q$ are complex coordinates of the branch points.  The branch cut connecting the branch points is a current sheet across which the magnetic field is discontinuous \cite{Green1965,Syrovatskii1971}

\begin{figure}[htp]
\psfig{file=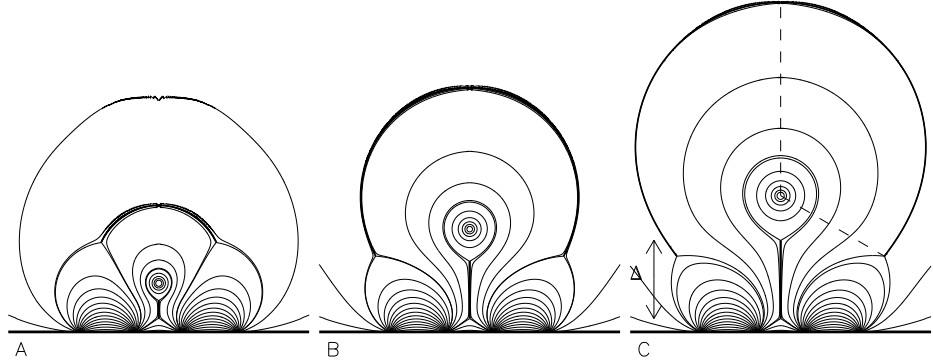,width=5.0in}
\caption{Equilibria with currents sheets occurring at each of the field's separators.
Equations  (\ref{eq:Q1})--(\ref{eq:Ah_eq}) are satisfied for specified $\theta$ and $\Delta$, and for
$A_h=1.4\,Q_1$ and $R=0.005\,x_1$.  The former parameters specify the size of each current 
sheet as illustrated in panel C.}
	\label{fig:AB}
\end{figure}

We consider a field where the lower null point, $y_b$, is replaced by a vertical current sheet  and the upper null point, $y_a$, an approximately horizontal sheet.  The vertical sheet extends between branch points $w=ip$ and $w=iq$ on the $y$ axis ($p<q$).  The right tip of the horizontal sheet is $w=\tau=r+is$ and the left tip is at $w=-\tau^*$.  Adding the image currents for 
each sheet \cite{Tur1976,Titov1992} gives a complex field
\be
  \hat{F}(w) ~=~ B_y+iB_x ~=~
  iD{\sqrt{(w^2+p^2)(w^2+q^2)}\sqrt{(w^2-\tau^2)(w^2-\tau^{*2})}\over
  (w^2-x_1^2)(w^2-x_2^2)(w^2+h^2)} ~~,
  	\label{eq:F}
\ee
which has replaced each of the four nulls in \eq\ (\ref{eq:F0}) with a branch cut.  The real coefficient $D$ is the dipole moment of the new field; at large distances 
$\hat{F}\sim iD/w^2$.  Examples of such a field are shown in \fig\ \ref{fig:AB} \cite[give other examples]{Tur1976,Hu1982}.

Each square root in \eq\ (\ref{eq:F}) has two branch points in the corona ($y>0$) between which runs a branch cut.  Across the branch cut the square root changes 
complex phase by $\pi$, and thus $\hat{F}\to-\hat{F}$.  This leaves the magnitude unchanged so $|\bvec|=|\hat{F}|$ is continuous, assuring pressure balance across the current sheet.  In order that the sheet is in force balance, \ie\ a tangential discontinuity, the branch cut must be parallel or antiparallel to the direction of $\hat{F}$ at every point.  This is simple to accomplish for the lower (vertical) current sheet, but more challenging for the upper sheet because of its curvature (Tur and Priest (1976), Aly and Amari (1989) and Titov (1992) all present versions a general methodology for doing so.)\nocite{Aly1989,Titov1992}

To see the structure of the vertical sheet we evaluate \eq\ (\ref{eq:F}) along the $y$ axis ($w=iy$) to find
\be
  \hat{F}(iy) ~=~ B_y + i B_x ~=~ iD{\sqrt{(y-p)(y-q)}
  \sqrt{(y+p)(y+q)}\,|y^2+\tau^2|\over
  (y^2+x_1^2)(y^2+x_2^2)(h^2-y^2)} ~~,
\ee
Inspection reveals that for points above or below the vertical current sheet, 
($y>q$ or $y<p$) both square roots are real and $B_y=0$.  To be consistent with the potential field we make the first square root positive above and negative below the sheet.  Between the branch points, $p<y<q$, the first square root is purely imaginary so $B_x=0$.  This means that taking the branch cut along the $y$ axis, between the branch points, satisfies the force-free condition.  This is the vertical current sheet.  

The branch is arranged so that $B_y>0$ on the $x>0$ side and $B_y<0$ on the other side.  The current density of the sheet is then found
\be
  J_z(x,y) ~=~ {1\over 4\pi}{\partial B_y\over \partial x} ~=~
  {1\over 2\pi}B_y(0^+,y)\delta(x) ~~,
\ee
where $\delta(x)$ is the Dirac delta function.  The surface density in the sheet is then $K(y)=|\hat{F}(iy)|/2\pi$ and the net current in the lower sheet is
\be
  I_b ~=~ {1\over 2\pi}\int_p^q|\hat{F}(iy)|\, dy ~~.
\ee

The upper current sheet is more complicated, but treated in roughly the same manner; details are provided in an appendix
(Aly and Amari (1989) and Titov (1992) present more rigorous derivations)\nocite{Aly1989,Titov1992}.  It is not, however, necessary to trace the path of the branch cut in order to compute the equilibrium.

There are six real parameters in \eq\ (\ref{eq:F}).  We express the upper branch point $\tau$ in terms of two real parameters, $\rho$ and $\theta$ defined by the relation
\be
   \tau ~=~ r+is ~=~ i(h+\rho e^{i\theta}) ~~.
\ee
These turn out to be better behaved than $r$ and $s$ in the solution of the equilibrium equations.  We also introduce a parameter defining the length of the lower current sheet, $\Delta=q-p$, illustrated in \fig\ \ref{fig:AB}. 

The six real parameters defining the equilibrium magnetic field are $D$, $h$, $p$, $\Delta$, $\rho$, and $\theta$.  Boundary conditions impose several constraints on these.  Matching the point sources at the photosphere requires
\begin{eqnarray}
  \res_{w=x_1} \hat{F} &=& 
  iD{\sqrt{(x_1^2+p^2)(x_1^2+q^2)}\sqrt{(x_1^2-r^2+s^2)^2 + 4 r^2s^2 }\over
  2x_1(x_1^2-x_2^2)(x_1^2+h^2)} ~=~ -iQ_1~~, \label{eq:Q1} \\
  \res_{w=x_2} \hat{F} &=& 
  iD{\sqrt{(x_2^2+p^2)(x_1^2+q^2)}
  \sqrt{(x_2^2-r^2+s^2)^2 +4 r^2s^2 }\over
  2x_2(x_2^2-x_1^2)(x_2^2+h^2)} ~=~ iQ_2 ~~.
  \label{eq:Q2}
\end{eqnarray}
One more condition is that the flux rope remain in force balance.  This means that the magnetic field at $y=h$ vanishes, after excluding the self-field,
\be
 \Byfr+i\Bxfr ~=~ {d\over dw}\left[(w-ih)\hat{F}(w)\right]_{w=ih} ~=~ 0 ~~.
 	\label{eq:B0}
\ee
Equations (\ref{eq:Q1}) -- (\ref{eq:B0}) are three independent constraints on the six parameters, leaving three free parameters.

One final constraint follows from the requirement that the flux between the photosphere and the flux rope maintain a fixed value
\be
  A_h ~+~ {\rm Re}\left[~\int\limits_0^{i(h-R)}\hat{F}(w)\, dw ~\right]~=~ 0 ~~,
  	\label{eq:Ah_eq}
\ee
where the integration path must be chosen not to cross the lower current sheet.
This, along with \eqs\  (\ref{eq:Q1}) -- (\ref{eq:B0}), provides four equations capable of fixing four of the unknowns.  We choose to vary $D$, $h$, $p$, and $\rho$, in order to satisfy the equations for specified values of the free
parameters $\theta$ and $\Delta$.

\subsection{Structure of Equilibrium Space}

Setting $\theta=\Delta=0$ means $q\to p$ and $-\tau^*\to\tau$ which shrinks the complex branch cuts to points thereby converting the current sheets back into X-points.  The solution should thus correspond to the X-point field treated in the previous section.  This has a single generically quadrupolar solution, $p=y_b$, $\rho=y_a-h$, $D=D_0$, and $h$ all taken from the solutions in \fig\ \ref{fig:Ah}.  When $\theta$ and $\Delta$ are set slightly greater than zero, the X-point solution can be used as an initial guess to solve \eqs\ (\ref{eq:Q1})--(\ref{eq:Ah_eq}) numerically using the Newton-Raphson algorithm.  Upon solution, $\theta$ and $\Delta$ can be increased again and a new solution found beginning with the prior values.  This is repeated to find equilibria for any values of $\theta$ and $\Delta$, such as those shown in \fig\ \ref{fig:AB}.

\begin{figure}[htp]
\psfig{file=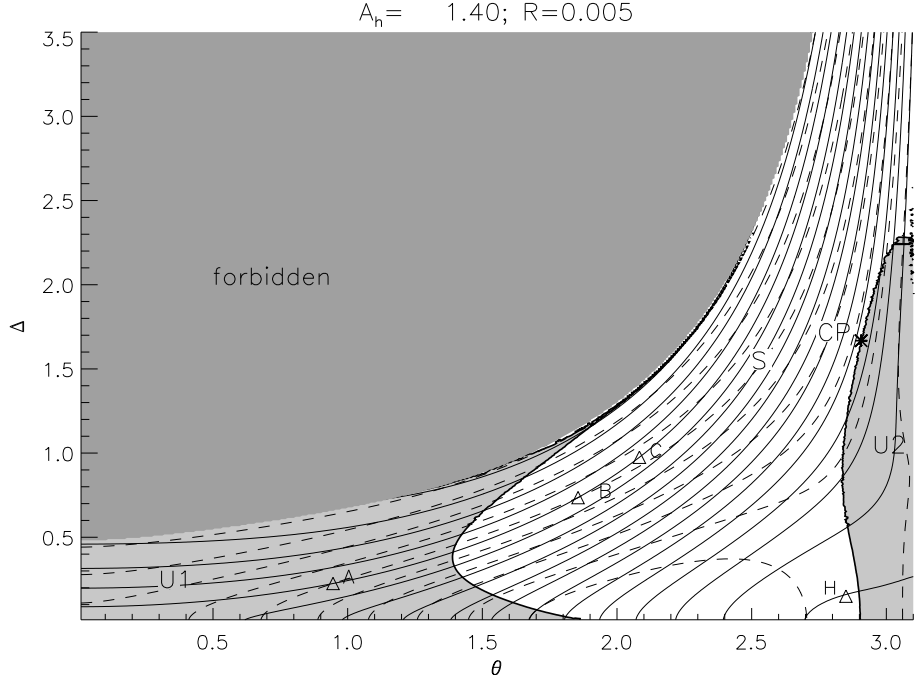,width=5.2in}
\caption{Contours of $\psi_1(\theta,\Delta)$ (solid) and $\psi_5(\theta,\Delta)$ (dashed) for equilibria with $A_h=1.4\,Q_1$ and $R=0.005\,x_1$.  Points labeled A and B have $\psi_1=0.5\,Q_1$ and $\psi_5=0.5\,Q_1$, and point C has
$\psi_1=0.5\,Q_1$ and $\psi_5=0.45\,Q_1$; these correspond to the equilibria shown in \fig\ \ref{fig:AB}.  The light grey regions labeled U1 and U2 contain unstable equilibria while the white region, labeled S, contains stable equilibria.  The darker grey region contains no equilibria, and is bounded by the locus of asymptotically octopolar equilibria with 
$\psi_5=0$.}
	\label{fig:psi15}
\end{figure}

For each pair $(\theta,\Delta)$, an equilibrium solution is computed as described above yielding a function, $\hat{F}(w)$  given in \eq\ (\ref{eq:F}).  This can be integrated from $w=0$ to $w=\tau$, or $w=ip$ to compute $A_a$ or $A_b$ respectively.  These can then be used to find the domain fluxes $\psi_1(\theta,\Delta)$ and 
$\psi_5(\theta,\Delta)$ as shown in \fig\ \ref{fig:psi15}.  Points where particular contours intersect correspond to an equilibrium with those values of $\psi_1$ and $\psi_5$. 

\begin{figure}[htp]
\centerline{\psfig{file=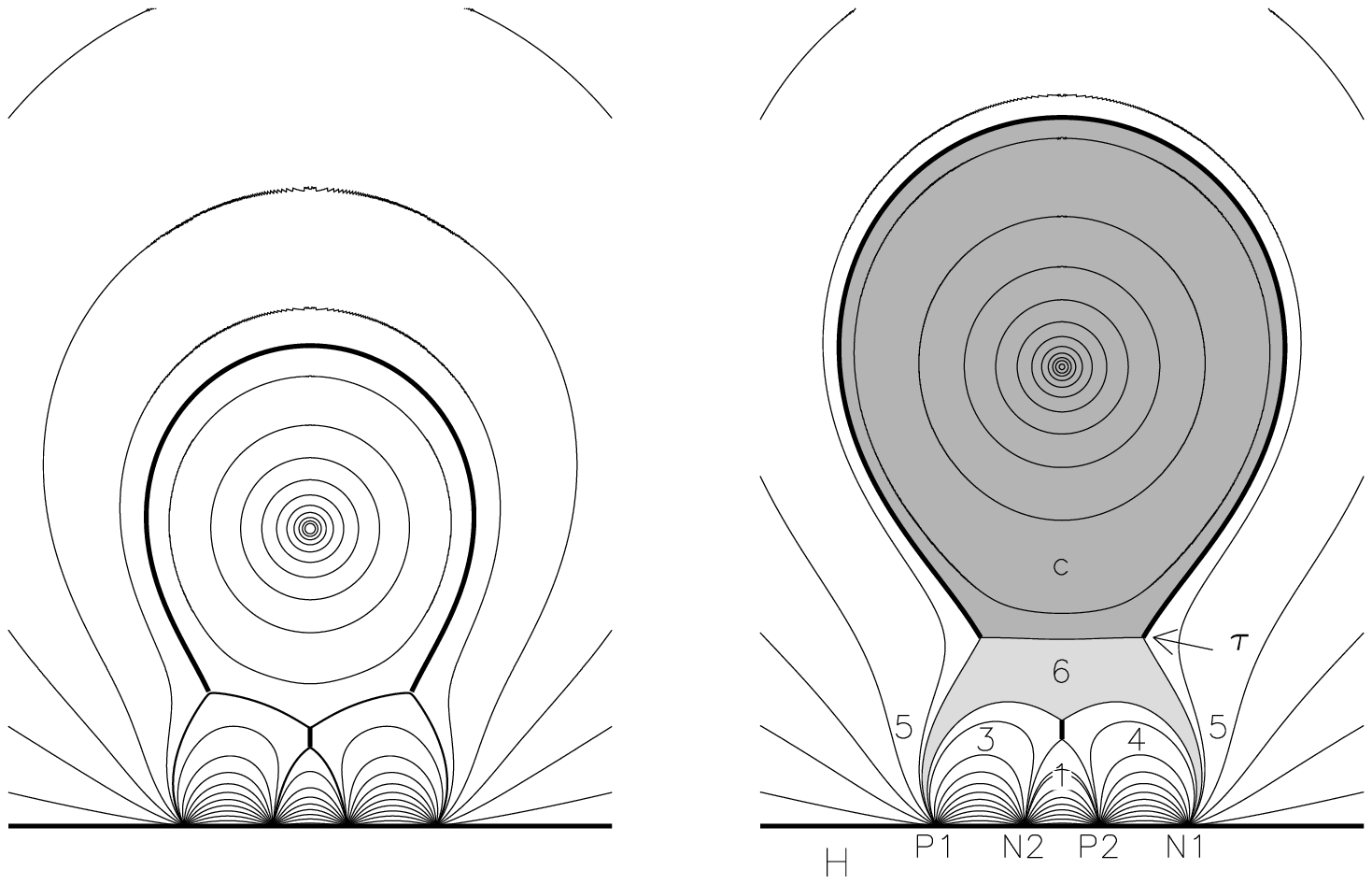,width=4.0in}~~\psfig{file=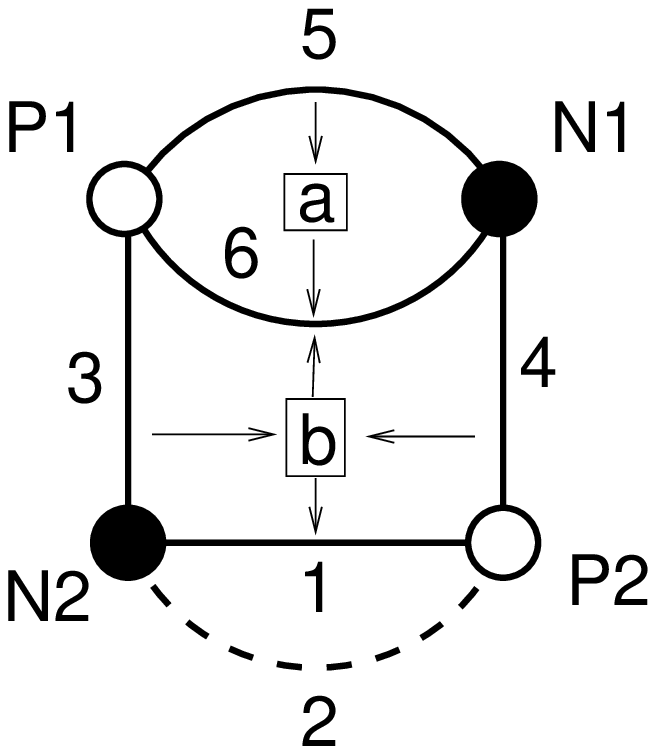,width=1.1in}}
\caption{An example of an untethered equilibrium (center) and the marginal state (left) intermediate between the tethered and untethered configurations.  The new domain, 6, is shaded with light grey, and the closed flux with dark grey.  All the domains are numbered.  The domain graph on the right shows the new domain and a dashed edge shows where domain 2 had been.}
	\label{fig:disc}
\end{figure}

There are, among our equilibria, fields of a distinct topology.  According to the analysis following \eqs\ (\ref{eq:graph}), the flux $\psi_2$ should equal
\be
  \psi_x ~=~ \psi_5~-~\psi_1 ~+~\pi(Q_2-Q_1) ~=~ A_b-A_a ~~,
\ee
assuming it is positive (all fluxes are unsigned quantities).  In cases where $\psi_x$ is negative domain 2 ceases to exists and is replaced by a new domain, labeled number 6, as shown in \fig\ \ref{fig:disc}.  For the new domain graph \eqs\ (\ref{eq:graph}) must be be replaced by
\be
  \psi_1+\psi_3 ~=~ \pi Q_2 ~~~~,~~~~
  \psi_3+\psi_5+\psi_6  ~=~ \pi Q_1 ~~,
  	\label{eq:graph2}
\ee
so $\psi_6=-\psi_x>0$.  The total flux under the flux rope is now $A_h=\psi_c+\psi_6-\psi_1$, so the closed flux surrounding  the flux rope is
\be
  \psi_c ~=~ A_h +\psi_1-\psi_6 ~~.
  	\label{eq:psic_unteth}
\ee

Because the inner bipole no longer has field lines passing over the flux rope (\ie\ domain 2), we refer to this new configuration as {\em untethered}, and to configurations including domain 2, such as \fig\ \ref{fig:A0}, as tethered.  It should be noted that, in general, both states include flux in domain 5 which does pass over the flux rope.
While the same current current sheets exist in both tethered and untethered states, their relation to the domains differs as illustrated in the graph on the right of \fig\ \ref{fig:disc}.  Aside from this the difference between the tethered and untethered states involves field line topology and is not necessarily a dynamical one.

A transition to this new kind of equilibrium can occur through a topological bifurcation involving an intermediate state where $A_a=A_b$, shown on the left of \fig\ \ref{fig:disc}.  In this case the separatrices from both current sheets coincide and there is neither a domain 2 nor a domain 6.  This is a two-dimensional version of the so-called {\em global spine-fan bifurcation} where a spine field line sweeps across a second separatrix \cite{Brown1999,Longcope2005b}; this bifurcation has been associated with break-out reconnection in three dimensions \cite{Maclean2005}.

The set of allowed equilibria is bounded by limiting cases.  As mentioned above, in the limits $\theta\to0$ or $\Delta\to0$ one of the current sheets (above and below respectively) reduces to an X-point.  It is possible to proceed beyond the $\Delta=0$ point, for example, by replacing, in \eq\ (\ref{eq:F}), the vertically separated branch points $ip$ and $iq$ with complex points, $\tau_p$ and $-\tau_p^*$, separated horizontally.  We do not consider this case here, and thus limit ourselves to the set $\theta\ge0$ and $\Delta\ge0$ as shown in \fig\ \ref{fig:psi15}.

The right edge of \fig\ \ref{fig:psi15} occurs where $\theta\to\pi$.  This is another case where distinct branch points converge to form an X-point.  In this case
\be
  \tau ~=~i(h+\rho e^{i\theta}) \to i(h-\rho) ~=~-\tau^* ~~.
\ee
Remarkably, the branch points converge {\em below} the flux rope leaving the entire branch cut (\ie\ the current sheet) surrounding it.  The resulting field,
\begin{eqnarray}
  \hat{F}^{(bc)}(w) &=& iD{\sqrt{w^2+p^2}\sqrt{w^2+\tilde{q}^2}\sqrt{w^2+t^2}\sqrt{w^2+u^2}\over
  (w^2-x_1^2)(w^2-x_2^2)(w^2+h^2)} \label{eq:Fbc} \\[7pt]
  &~&~~~\hbox{where}~p<\tilde{q}<t<u<h \nonumber 
\end{eqnarray}
has a vertical current sheet extending between $w=ip$ and $w=iu$, with a gap between $w=i\tilde{q}$ and
$w=it$; $D$ is the dipole moment of this new field.  One vertical branch cut, extending between $w=ip$ and $w=i\tilde{q}$, forms the lower section of the current sheet with $I_b>0$.  The branch cut from $w=it$ extends vertically upward to $w=iu$.  It then wraps entirely around the current filament $w=ih$ before ending at the same branch point ($w=iu$).  This is the upper current sheet, with $I_a<0$, which completely encloses the current filament  as well as extending part of the way along the vertical beneath the flux rope.  The flux passing through the gap, $\tilde{q}<y<t$, of this 
{\em broken current-sheet} composes domain 6, so field represented by \eq\ (\ref{eq:Fbc}) is an untethered field.

\begin{figure}[htp]
\psfig{file=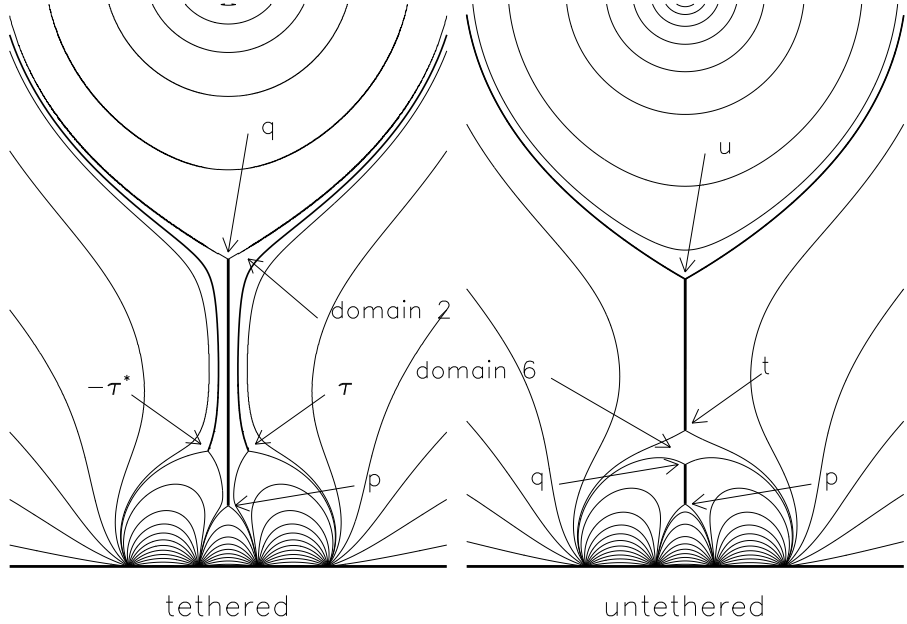,width=5.2in}
\caption{Transition from tethered (left) to untethered (right).  Both equilibria have $\psi_1=1.7Q_1$.  In the tethered equilibrium, on the left $\psi_2=0.014Q_1$, while in the untethered equilibrium on the right $\psi_6=0.018Q_1$.  These domains are indicated by arrows as are all four branch points in each case.}
	\label{fig:gap}
\end{figure}

The transition to untethered geometry of \eq\ (\ref{eq:Fbc}) occurs as $\theta\to\pi$, so $\tau$ and $-\tau^*$ both approach the imaginary axis.  This can occur above the lower current sheet so $\tau\to it=iu$ and then $q\to\tilde{q}$.  It can also occur within the lower current sheet to create a gap and reverse the upper portion, $\tau\to i\tilde{q}=it$ and $q\to u$, 
depicted in \fig\ \ref{fig:gap}.  In the tethered version (left) the top portion of the lower (vertical and positive) current sheet is sandwiched between vertical sections of the upper current sheet (negative).  Upon transition these three sections merge into a single vertical segment of the negative current sheet: $t<y<u$.

Equilibria of the form given by \eq\ (\ref{eq:Fbc}) are not represented within the $(\theta,\Delta)$ space of \fig\ \ref{fig:psi15}.  They form a different equilibrium space which joins along the right edge ($\theta=\pi$).  Thus we cannot consider that edge to be a limit on the space.  Broken-sheet equilibria like that on the right of \fig\ \ref{fig:gap}, do not, however, resemble the stressed pre-eruption fields found in numerical simulations.  We will therefore continue to focus our analysis on the $(\theta,\Delta)$ space of \fig\ \ref{fig:psi15}, and return to consider the broken-sheet case when discussing eruption.

\begin{figure}[htp]
\centerline{\psfig{file=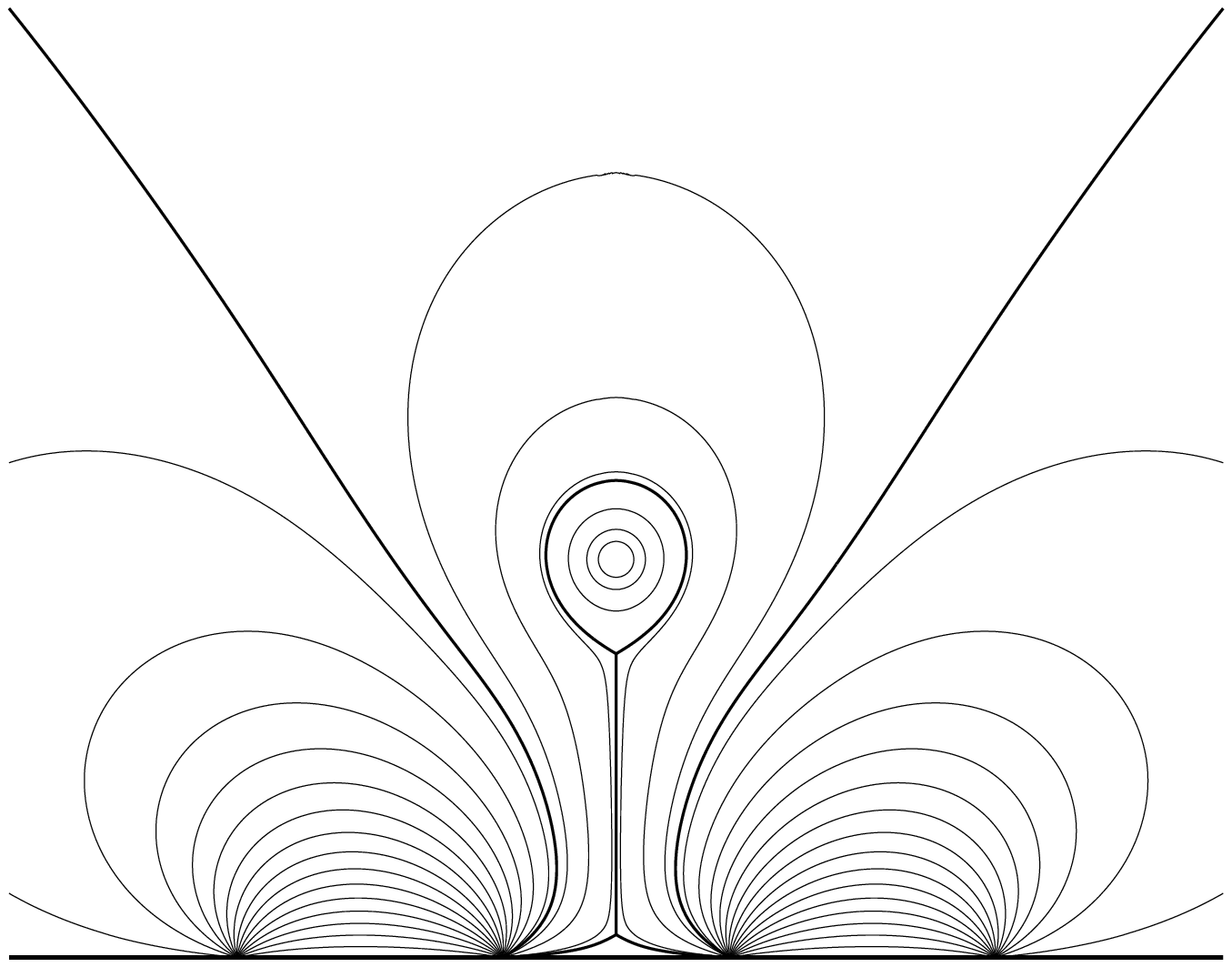,width=4.0in}\psfig{file=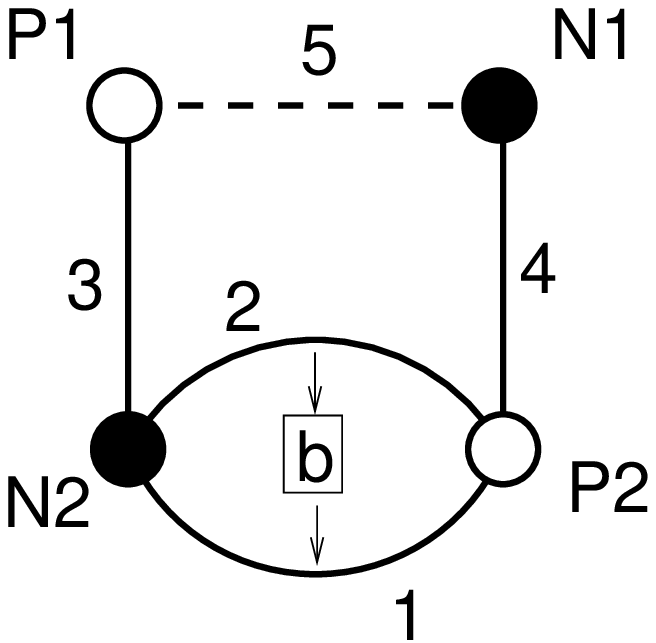,width=1.1in}}
\caption{The asymptotically octopolar field $\hat{F}^{(ao)}(w)$ from \eq\ (\ref{eq:Fo}).  The domain graph is shown on the right with a dashed edge indicating the domain $\psi_5$ which no longer exists.}
	\label{fig:as_oct}
\end{figure}

Another limiting case is the equilibrium with $\psi_5=0$, shown in \fig\ \ref{fig:as_oct}.  This field
generated by a limit of \eq\ (\ref{eq:F}) in which $\tau\to\infty$ and $D|\tau|^2\to M$, a constant, 
\be
  \hat{F}^{(ao)}(w) ~=~
  iM{\sqrt{(w^2+p^2)(w^2+q^2)}\over
  (w^2-x_1^2)(w^2-x_2^2)(w^2+h^2)} ~~.
  	\label{eq:Fo}
\ee
This has only one branch cut corresponding to the lower current sheet, since the upper current sheet has moved to infinity.  The function asymptotically approaches $\hat{F}(w)\to iM/w^4$; it is {\em asymptotically octopolar}.  The asymptotic flux function
\be
  A^{(ao)}(r,\phi) ~\to~ {M\sin(3\phi)\over 3r^3} ~+~ \pi(Q_1-Q_2) ~~,
\ee
has separatrices at $\phi=\pi/3$ and $\phi=2\pi/3$, separating domains 2, 3, and 4.  Domain 5 no longer exists.    It is evident from the graph that in the asymptotically octopolar state $\psi_3=\pi Q_1$ and $\psi_1+\psi_2=\pi(Q_2-Q_1)$.  These equilibria separate regions U1 and S from the forbidden region in \fig\ \ref{fig:psi15}.

It is not possible to conclude definitively that specified values of $(\theta,\Delta)$ admit only one equilibrium solution, the one used in \fig\ \ref{fig:psi15}.  Those equilibria compose only the set continuously related to the unique X-point field, in the manner described above.   However, since the contours of $\psi_1(\theta,\Delta)$ and  $\psi_5(\theta,\Delta)$ are smooth and non-singular, we see no point at which our equilibrium manifold might join a second.  We show below that the same is not the case when equilibria are designated by $\psi_1$ and $\psi_5$.  Contours of $\theta(\psi_1,\psi_5)$ and 
$\Delta(\psi_1,\psi_5)$ do exhibit singular behavior and those equilibria are {\em not} unique.  The absence of a similar criticality in \fig\ \ref{fig:psi15} suggests that equilibria are uniquely identified by the parameters 
$\theta$ and $\Delta$, at least up to the $\theta=\pi$ edge.

\subsection{Stability of Equilibria}

We now consider the stability of the equilibria to ideal vertical displacement of the current filament.  By ideal we mean evolution which preserves $A_h$ as well as all other domain fluxes; these are equivalent to preserving $\psi_1$ and $\psi_5$.  The stability question is intimately related to the uniqueness of equilibria corresponding to a specified pair $(\psi_1,\psi_5)$.  In other words we wish to invert the mapping $(\theta,\Delta)\mapsto(\psi_1,\psi_5)$ shown through contours in \fig\ \ref{fig:psi15}.  Since there are some contours which cross each other more than once, as at the points labeled
 A and B, the inverse mapping will not be single-valued.  This means that a given pair of flux values corresponds to multiple possible equilibria.  As an illustration, the left and center fields in \fig\ \ref{fig:AB}, both have 
 $\psi_1=\psi_5=0.5Q_1$, and thus have all the same domain fluxes.  It is therefore possible to convert one to the other through purely ideal motions.
 
Cases of two intersections between contours are separated from cases of no intersection by the case of contour 
{\em tangency}
\be
  {\partial\psi_1\over\partial \theta}{\partial\psi_5\over\partial \Delta} ~-~
  {\partial\psi_1\over\partial\Delta}{\partial\psi_5\over\partial \theta} ~\equiv~ {\partial(\psi_1,\psi_5)\over
  \partial(\theta,\Delta)} ~=~ 0 ~~,
  	\label{eq:tangency}
\ee
where the notation $\partial(x_1,x_2)/\partial(y_1,y_2)$ refers to the determinant of the $2\times2$ Jacobian matrix 
$\partial x_i/\partial y_j$.  This condition defines two curves shown in \fig\ \ref{fig:psi15} as a dark solid curve bounding light grey regions U1 and U2.  In each case where a contour intersects twice it does so on opposite side of the tangency curve, as in the case of points A and B.  There is a single point, called a {\em critical point} (CP), on the right tangency curve (bordering U2) where the curve itself is tangent to the contours of both $\psi_1$ and $\psi_5$.  This marks the point  where both $\psi_1$ and $\psi_5$ assume their minimum value along the curve.  Contours passing into region U2 can therefore intersect three times: once inside U2 and twice more in S.

\begin{figure}[htp]
\psfig{file=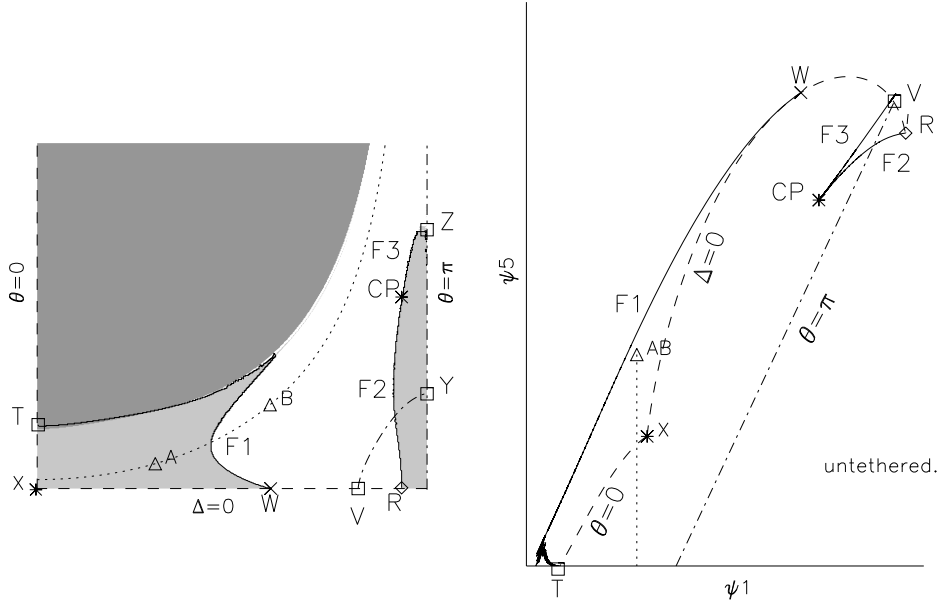,width=5.2in}
\caption{The mapping of equilibrium space $(\theta,\Delta)$ (left) onto flux space $(\psi_1,\psi_5)$ (right).
Points labeled T, X, W, etc. map onto each other.  The two triangles A and B, match the points 
in \fig\ \ref{fig:psi15} and lie on the dotted contour $\psi_1(\theta,\Delta)=0.5Q_1$.  Both points 
map onto the single triangle labeled AB, and the contour maps to the vertical dotted line.
The mapping folds along the dark curves labeled F1 and F2 matching the curve 
from \eq\ (\ref{eq:tangency}).  Broken curves, terminating in point V, denote $\psi_2=0$ and therefore
separate tethered from untethered states.}
	\label{fig:map}
\end{figure}

A mapping from one plane, in this case $(\theta,\Delta)$, onto a second, $(\psi_1,\psi_5)$, can become non-invertible in a limited number of ways.  The present mapping involves a ``folding'' of the surface, known as a 
{\em cusp catastrophe}, which produces a multi-valued inverse \cite{Poston1978}.   The mapping here, illustrated 
in \fig\ \ref{fig:map}, involves folds along the tangency curves, \eq\ (\ref{eq:tangency}), labelled F1, F2, and F3.  The grey region U1 is folded over and mapped onto the region bounded by F1 and the dashed curve T--X--W on the right panel of \fig\ \ref{fig:map}.  Thus the point A lies on the upper fold over top the point B --- the two occupy a single point in 
$(\psi_1,\psi_5)$ space, labeled AB on the right panel.  The region U2 is folded back on itself with two folds, F2 and F3, which meet at the critical point CP.  The wedge of $(\psi_1,\psi_5)$ space between these curves maps back to the {\em three} points, two in S and one in U2.

The folded mapping produces a multi-valued inverse mapping, and thus leads to multiple equilibria for specified domain fluxes.   This is a multidimensional analog of the situation in Section \ref{sec:stabX}, where the inverse of the curve $A_h(h)$ is multivalued leading to equilibria QL and QH shown in \fig\ \ref{fig:Ah}.  As in that case, we expect the pairs of equilibria to be of opposing stability: one unstable and one stable.  This is a standard scenario in cases of cusp catastrophes \cite{Forbes1991}.

Following the analysis of Section \ref{sec:stabX}, instability to ideal motions occurs when
\be
  I_0\,\left.{\partial \Bxfr\over \partial h}\right)_{\psi_1,\psi_5,A_h} ~=~
   I_0\,\displaystyle{\partial(\psi_1,\psi_5)\over\partial(\theta,\Delta)}
  \left[\displaystyle{\partial(\psi_1,\psi_5,h)\over\partial(\theta,\Delta,\Bxfr)}\right]^{-1} ~>~0 ~~,
  	\label{eq:unstab}
\ee
where all partial derivatives hold fixed two of the three unknowns $\theta$, $\Delta$ or $\Bxfr$.  
This quantity will vanish along the tangency curve, \ie\ \eq\ (\ref{eq:tangency}), so that is also a curve across which stability changes.  It is a characteristic of such cusp catastrophes that of two simultaneously permitted equilibria one will be stable and one unstable.

\begin{figure}[htp]
\psfig{file=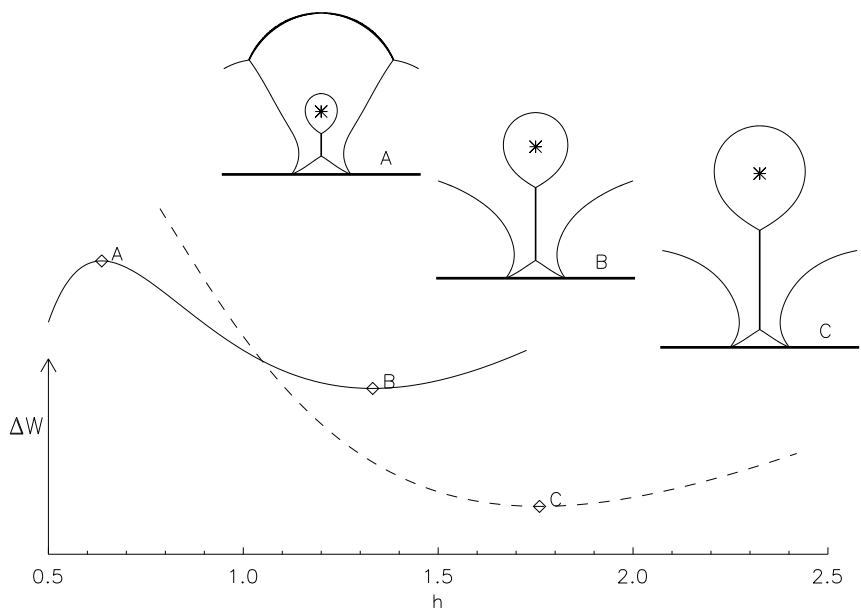,width=5.1in}
\caption{The work computed by integrating $F_y$ over $h$ at constant $\psi_1$ and $\psi_5$.  The solid curve, for 
$\psi_1=0.5Q_1$ and $\psi_5=0.5Q_1$, includes equilibria A and B from \fig\ \ref{fig:psi15}.  The dashed curve 
is for $\psi_1=0.5Q_1$ and $\psi_5=0.45Q_1$.  The diagram along the top show separatrices (thin curve), current sheets (dark curve), and flux ropes (asterisk), for equilibria A, B, and C.}
	\label{fig:work}
\end{figure}

To test our interpretation of the stability boundary we begin from one equilibrium, say equilibrium A, and find related fields $\hat{F}$ with different values of $h$.  Values of $D$, $p$, $\rho$, $\theta$, and $\Delta$ are all varied to satisfy 
\eqs\ (\ref{eq:Q1}), (\ref{eq:Q2}), (\ref{eq:Ah_eq}), and constraints on $\psi_1$ and $\psi_5$.  The flux rope current
\be
  I_0 ~=~\half \res_{w=ih} \hat{F} ~=~ 
  D{\sqrt{(h^2-p^2)(h^2-q^2)}  \sqrt{(h^2+r^2-s^2)^2 +4 r^2s^2 }
  \over  4h(h^2+x_1^2)(h^2+x_2^2)} ~~,
\ee
is evaluated and used to compute the $F_y(h)=I_0\Bxfr$.  This is evaluated over a range of heights and integrated to compute work.  The result, shown in \fig\ \ref{fig:work}, includes two equilibria where 
$F_y=-\partial(\Delta W)/\partial h=0$, corresponding to points A and B.  These are the two values of $h$ where \eq\ (\ref{eq:B0}) is also satisfied.  We can repeat the computation above for other values of the fluxes $\psi_1$ and $\psi_5$.  If we keep $\psi_1=0.5Q_1$, the same as for equilibria A and B, but decrease $\psi_5$ from $0.5Q_1$ to $0.45 Q_1$ we obtain the dashed curve in  \fig\ \ref{fig:work}.  This has a stable equilibrium C with lower energy than B.

Based on the above analysis we conclude that equilibrium A is unstable and equilibrium B, on the other side of the tangency curve, is stable.  We infer that all equilibria in region U1 are unstable while those in region S are stable.  The unstable region U1 includes the X-point equilibrium, X.  This is consistent with the conclusion we reached in Section \ref{sec:stabX} even without accounting for the conservation of fluxes $\psi_1$ and $\psi_5$.  The region between F2 and F3 maps to three equilibria, of which one is unstable (in U2) and is sandwiched between two stable equilibria from S.

The region in \fig\ \ref{fig:map}  to the lower right of the broken curve V--Y, consists of untethered equilibria.  These include both stable and unstable varieties separated by F2.  The field's topology thus does not determine its stability: tethered and untethered fields come in both stable and unstable varieties.

\begin{figure}[htp]
\psfig{file=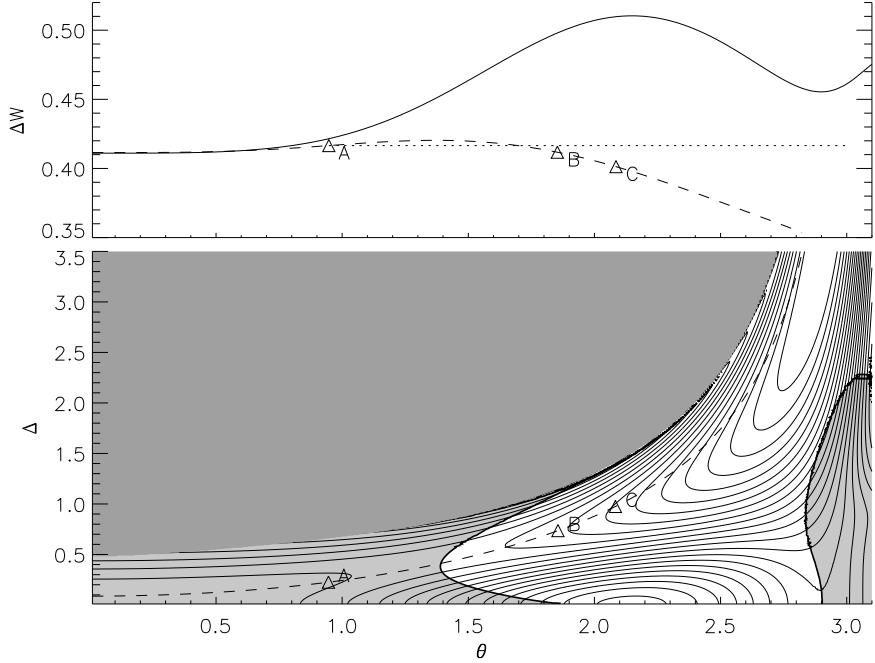,width=5.1in}
\caption{The free energy $\Delta W(\theta,\Delta)$ of equilibria.  The bottom panel shows contours of the function within the same space as \fig\ \ref{fig:psi15}.  Light grey regions are U1 and U2.  The dashed contour shows 
the curve $\psi_1=0.5Q_1$, along which equilibria A, B, and C fall.  The upper panel shows $\Delta W$ along the $\Delta=0$ line (solid) and the $\psi_1=0.5Q_1$ contour (dashed).}
	\label{fig:erg}
\end{figure}

An alternative to the work integral is the magnetic free energy 
\begin{eqnarray}
  \Delta W 
  &=& {1\over 8\pi}\int\nabla (A+ A_0)\cdot
  \nabla (A-A_0)dx\,dy \nonumber \\
  &=& \half I_0 [A_h-A_0(h)] ~+~
  \sum_{j=a,b}{1\over 4\pi} \int\limits_{{\cal C}_j} |B|\,(A-A_0)\, d\ell,
\end{eqnarray}
where ${\cal C}_a$ and  ${\cal C}_b$ are the branch cuts defining upper and lower current sheets.  This is shown in \fig\ \ref{fig:erg}.  Both tangency curves cross saddle points in the free energy.  This is understandable because ideal motions can move the state along contours of $\psi_1$ and $\psi_5$ where they are tangent.  Since every point is an equilibrium, the energy must be unchanged by this allowed motion.

The free magnetic energy can be alternatively derived by integrating the electromagnetic work required to change the flux function at each current sheet to its final value
\be
  \Delta W ~=~\int(\, I_a\, dA_a ~+~ I_b\, dA_b \,) ~~.
  	\label{eq:erg_I}
\ee
The free energy will not depend on the order in which the fluxes are changed provided that
\be
  {\partial A_a\over \partial I_b} ~=~{\partial A_b\over \partial I_a} ~~;
\ee
which is a property of the equilibria in \fig\ \ref{fig:psi15}.  These two derivatives are the differential mutual inductances, ${\cal L}_{ab}$ and ${\cal L}_{ba}$, which must be equal for the magnetic energy to the independent of sequence.  Moreover, the determinant of the differential inductance matrix
\be
  {\rm det}({\cal L}) ~=~ {\cal L}_{aa}{\cal L}_{bb}-{\cal L}_{ab}{\cal L}_{ba} ~=~
  {\partial(\psi_5,\psi_1)\over\partial(\theta,\Delta)}\,\left[ {\partial(I_a,I_b)\over\partial(\theta,\Delta)}\right]^{-1} ~~,
\ee
changes sign at the tangency boundary, like the force derivative, \eq\ (\ref{eq:unstab}).  The unstable equilibria share the property that the differential mutual inductances between the current sheets conspire to allow current increases to decrease the free magnetic energy.

\section{System Evolution}
\subsection{Catastrophic Instability Triggered by Reconnection}

The fields described above can evolve in two distinct ways.  Ideal evolution is driven by changes to the photospheric flux distribution, $x_j$, $Q_j$, or to the current in the filament through changes in $A_h$. 
Subduction occurs through a combined decrease in $Q_2$ and $A_h$.  The alternative, reconnection, is driven by electric fields at the coronal current sheets.  This form of evolution has been the focus of much recent research 
\cite{Antiochos1999,Moore2001}, and will be our focus here.

Magnetic reconnection is a complicated process which is still understood only incompletely.  In the end, it occurs because an electric field of some kind occurs within a current sheet.  This transfers flux out of two flux domains adjacent to the sheet's sides into domains at its tips.  By changing the global flux distribution, the electric field moves the system to a state with a lower equilibrium magnetic energy: it {\em releases} free magnetic energy.  The reconnection electric field itself may dissipate some energy, although due to the small volume it occupies this will not be comparable to the free energy of the global system \cite{Longcope2004d,Longcope2011b}.  The plasma flows accompanying the flux transfer are expected to be comparable to the Alfv\'en speed and therefore to account for significant energy.  Some fraction of this will be thermalized \cite{Longcope2009}, some fraction will radiate away as fast magnetosonic waves \cite{Longcope2012}, and some may persist as turbulence or trapped Alfv\'en waves, losing energy subsequently to non-thermal particles \cite{Miller1997}.  In the end we expect these processes to permit the system to achieve its new magnetostatic equilibrium, with a lower free energy.

Here we will focus on the net energy released by magnetic reconnection due to the flux transfer alone; we will not concern ourselves with the mechanism responsible for initiating or maintaing the electric field, or the processes responsible for thermalizing and radiating that energy.  We will simply assume that at some point during quasi-static evolution a reconnection electric field, $E_z$, does occur in one or both the equilibrium current sheets.  This will transfer flux at a rate 
$d\psi/dt=E_z$ which we assume is small enough for the evolution it induces to be treated quasi-statically.

Magnetic reconnection at the current sheets will change the flux function $A_a$ or $A_b$.  This will change the energy, according to \eq\ (\ref{eq:erg_I}),
\be
  {d(\Delta W)\over dt} ~=~ I_a{d A_a\over dt} + I_b{d A_b\over dt} ~=~
  -I_a{d\psi_5\over dt} ~-~ I_b{d\psi_1\over dt} ~~.
\ee
The sense of change depends on the relative sense of the signs of the electric field ($E_j=-dA_j/dt$) and current.  Energetically favorable reconnection always transfers flux from the domains adjacent to the sheet's sides into domains at its tips.  This energetically favorable sense is indicated on the domain graphs of \figs\ \ref{fig:A0}, \ref{fig:turtle}, \ref{fig:disc}, and \ref{fig:as_oct}.  

\begin{figure}[htp]
\psfig{file=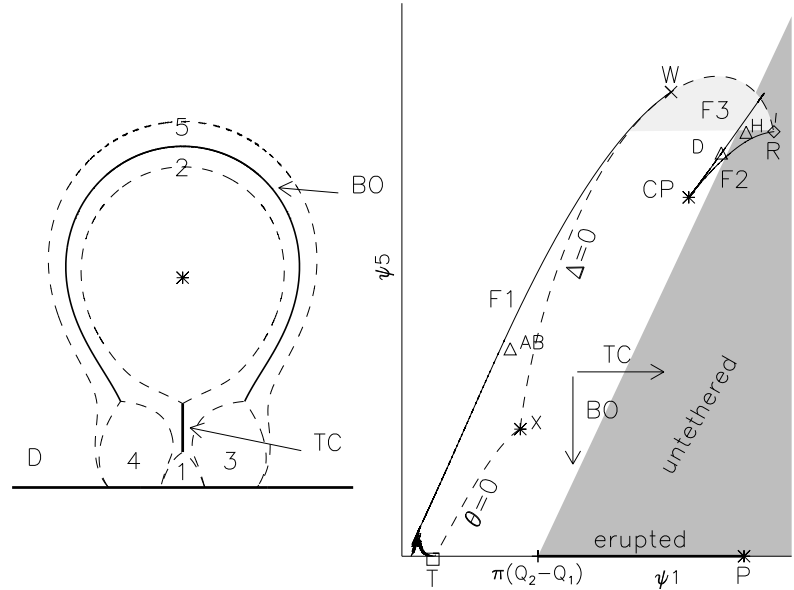,width=5.1in}
\caption{The effect of reconnection on position in equilibrium space $(\psi_1,\psi_5)$.  The structure of the space is shown on the right in the same manner as \fig\ \ref{fig:map}.  Arrows indicate the motion produced by breakout (BO) and tether-cutting (TC) reconnection.  The locations of the reconnection is indicated, on the left, using equilibrium D with 
$\psi_1=1.35Q_1$ and $\psi_5=0.9Q_1$, shown on the left.  The corresponding point is indicated with a triangle on the right. }
	\label{fig:rx_phase}
\end{figure}

Current $I_a$ in the upper sheet is negative ($I_a<0$, directed into the page) and reconnection there, which we refer to as {\em breakout reconnection} following Moore and Sterling (2006)\nocite{Moore2006}, decreases $\psi_5$.  This removes flux from domains 2 and 5 and adds flux to side lobes 3 and 4.  Current $I_b$ in the lower sheet is positive ($I_b>0$) and reconnection there, referred to as {\em tether-cutting reconnection}, increases $\psi_1$.  It 
removes flux from domain 2 over the flux rope and adds it domain 1 above the PIL as well as adding to the closed flux around the rope.  The effect of these reconnection processes in $(\psi_1,\psi_5)$ space is illustrated in \fig\ 
\ref{fig:rx_phase}.

\begin{figure}[htp]
\psfig{file=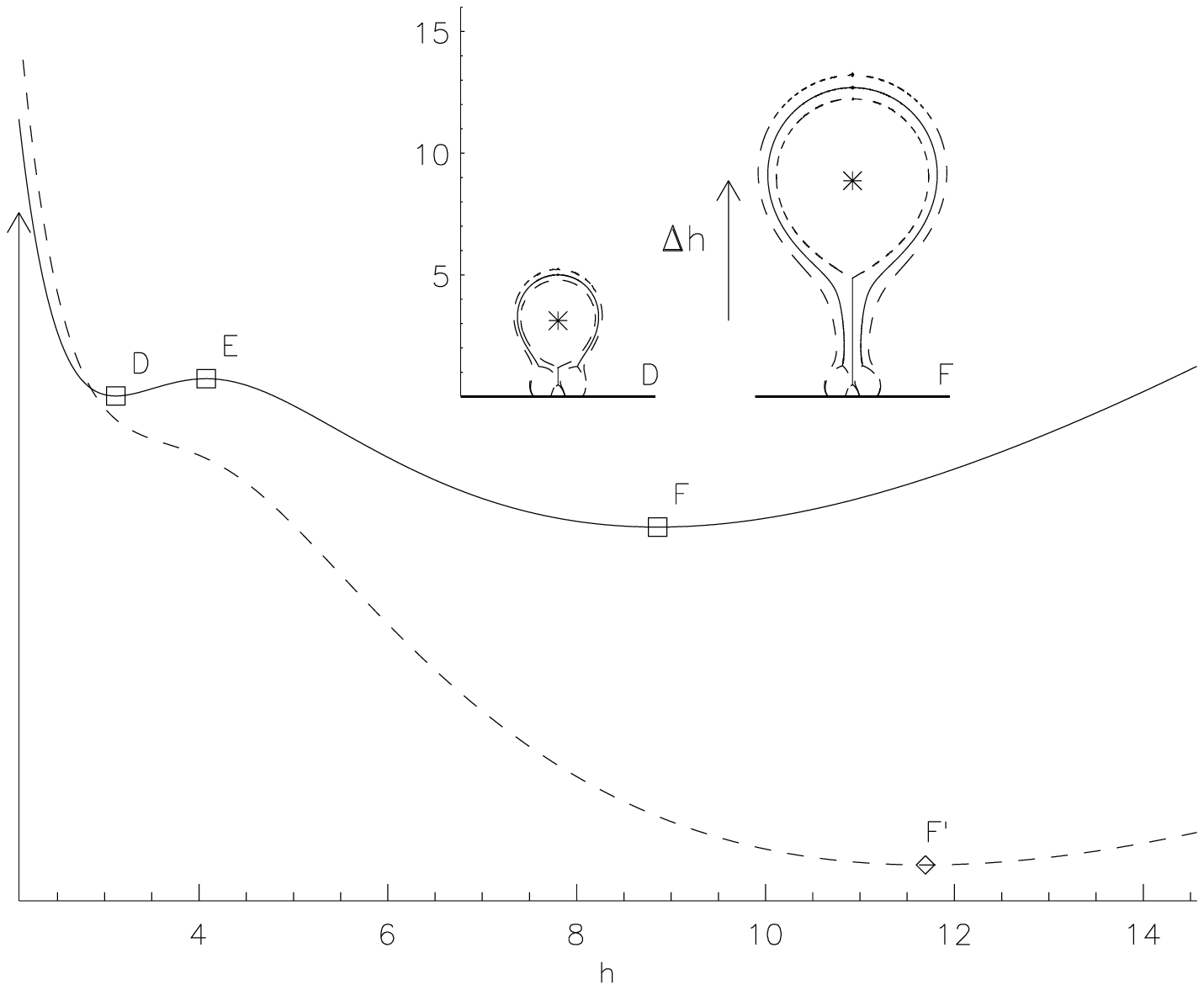,width=5.1in}
\caption{Force $F_y(h)$ integrated for vertical displacement from equilibrium D: $\psi_1=1.475Q_1$, $\psi_5=0.925Q_1$.  The solid curve shows the energy found from this integral, and D, E, and F are the equilibria.  The dashed curve is for the integral with $\psi_5=0.91Q_1$. Only one equilibrium F' exists in this case.}
	\label{fig:loe}
\end{figure}

By tracking the system through a space of equilibria we are tacitly assuming it evolves slower than dynamical time scales --- it is quasi-static evolution.  This becomes untenable when the evolutionary track encounters a fold in the inverse mapping, such as F1 or F2.  This is illustrated by the case of equilibrium D on \fig\ \ref{fig:rx_phase}.  This is a stable equilibrium, but either kind of reconnection will drive it onto fold F2.  Viewed in $(\theta,\Delta)$ space, in \fig\ \ref{fig:psi15}, F2 separates stable equilibria in region S from unstable equilibria U2.  As the evolving state encounters the fold the two equilibria merge and disappear in a {\em saddle-node bifurcation}, as illustrated in \fig\ \ref{fig:loe}; this is known as {\em loss of equilibrium}.  Prior to the bifurcation there was a second stable equilibrium, F, located in the stable region S across the second fold, F3.  When equilibria D and E undergo their saddle-node bifurcation the only equilibrium remaining is F' on the dashed curve of \fig\ \ref{fig:loe}.

When the loss of equilibrium occurs the system must change, on dynamical time scales, to assume the new equilibrium with a much higher flux rope position: $h$ increases from $3.5x_1$ to $9.0x_1$ between equilibria D and F, and $10.7x_1$ at F'.  This sudden dynamical evolution cannot be considered quasi-static.  Indeed, the inertia of the flux rope will naturally carry it beyond its equilibrium position \cite{Forbes1991}.  It will then oscillate about the equilibrium until some dissipative process can bring it to equilibrium.

From its position in \fig\ \ref{fig:rx_phase}, it is evident that equilibrium D can be brought to fold F2 by either breakout reconnection (moving downward) or tether-cutting reconnection (moving rightward).  The contours in \fig\ \ref{fig:psi15} shows that either following the solid contours in the direction down the dashed contours or following dashed contours down the solid contours, can lead to fold F2 (below the critical point CP).  Motions of this kind lead {\em away} from fold F3, so this will not be involved in a loss of equilibrium.  For the same reason F1 will not lead to loss equilibrium.  Once triggered, the dynamical evolution will carry the system across the fold F3 to a new equilibrium in S.

All three layers of $(\theta,\Delta)$ converge at the critical point CP.  This means that the three different equilibria converge to that same point.  At this point $dF_y/dh=0$ so the work function is quartic rather than quadratic, and thus has a vanishing frequency of oscillation.  Evolution through this point is not, strictly speaking, a loss of equilibrium.  It does, however, violate the quasi-static assumption since the dynamical time-scale has become infinite.  Reconnection at the upper current sheet, for example, will produce a quasi-static change in the flux rope height
\be
  {dh\over dt} ~=~ {d\psi_5\over dt} \,\left.{\partial h\over\partial \psi_5}\right)_{\psi_1,\Bxfr}
  ~=~ E_a\,{\partial(h,\psi_1)\over\partial(\theta,\Delta)}
  \left[{\partial(\psi_1,\psi_5)\over\partial(\theta,\Delta)}\right]^{-1}~~,
  	\label{eq:dhdt}
\ee
where $E_a$ is the electric field inside the upper current sheet.  This diverges at the tangency curve, so the evolution cannot be treated quasi-statically.  Notably, we consider reconnection to be slow if $E_a$ is small, but here we see that an arbitrarily small electric field is capable of producing rapid changes.

For those equilibria to the left of the critical point in \fig\ \ref{fig:rx_phase} (\ie\ $\psi_1<1.32Q_1$) breakout reconnection will not trigger a loss of equilibrium.  Equilibria below the critical point ($\psi_5<0.82Q_1$) or within the light grey region 
in \fig\ \ref{fig:rx_phase} ($\psi_5>0.97Q_1$) cannot be brought to loss of equilibrium by tether-cutting reconnection.  In all of these cases the evolution can proceed quasi-statically.  Remaining quasi-static will, however, become increasingly difficult as small changes in flux result in large changes in flux rope height.  While the derivative $\partial h/\partial\psi_5$ in \eq\ (\ref{eq:dhdt}) does not diverge, it becomes extremely large as $\Delta$ becomes large.  It will be large even for small values of $E_a$ --- \ie\ even when reconnection is slow.

The reconnection-driven evolution assumes an unusual nature in the regions $\theta\to\pi$ or large $\Delta$, suggested by the upward sloping contours in the upper right of \fig\ \ref{fig:psi15}.  As a reconnection electric field transfers flux across a sheet, say the lower sheet, the sheet becomes longer, \ie\ $\Delta$ increases.  It is not evident from \fig\ \ref{fig:psi15} but the current in the sheet, $I_b$, increases as well.  This behavior is opposite to the more conventional Syrovatskii-Green current sheets \cite{Green1965,Syrovatskii1971} where flux transfer reduces the length of the sheet and its current.  This behavior is found in two and three-dimensional separators generally provided they carry small enough current, remain at a fixed location, and are not strongly coupled to other currents through mutual inductance 
\cite{Longcope1996d,Longcope2004}.  Its anomalous behavior indicates our present quadruploar case violates one or more of these conditions.

\subsection{Eruption and the Aly-Sturrock Conjecture}

Configuration D in \fig\ \ref{fig:loe} can be driven by breakout reconnection to suffer a loss of equilibrium.  This is demonstrated by the reduction of $\psi_5$ changing the work function from the solid to the dashed curve.  The height of the flux rope then changes suddenly, and eventually settles into the value for F', $h\approx 10.7x_1$.  As sudden as it is, this does not represent a complete eruption since the flux rope remains at a finite height.  It is evident from the dashed curve in \fig\ \ref{fig:loe} that the free energy increases with height beyond that particular equilibrium.

The stability of equilibria F and F' can be understood by considering the behavior of the field as we take  
$h\to\infty$ while holding constant $\psi_1$ and $\psi_5$.  This means $\psi_2>0$, and we must preserve a current sheet anchored to branch points at $w=\tau$ and $w=-\tau^*$ to separate domains 2 and 5.  These branch points will move, but must remain finite even as we take $h\to\infty$, since their separatrices overlie domains 3 and 4.  The branch point at 
$w=iq$ will, however, follow the flux rope at $w=ih$ as it diverges.  We therefore take the limit $h\to\infty$ and $q\to\infty$ in 
\eq\ (\ref{eq:F}) to find the {\em partially open} field
\be
  \hat{F}^{(po)} ~=~ i\tilde{Q}{\sqrt{w^2+p^2}\sqrt{w^2-\tau^2}\sqrt{w^2-\tau^{*2}}\over(w^2-x_1^2)(w^2-x_2^2)} ~~,
  	\label{eq:Fpo}
\ee
where $Dq/h^2\to \tilde{Q}$ is finite in this limit. This field is asymptotically monopolar, 
$\hat{F}^{(po)}(w)\to i\tilde{Q}/w$, and thus has infinite energy.  It is for this reason that both the solid and dashed work curves turn upward in \fig\ \ref{fig:loe}: they must go to infinity as $h\to\infty$.

This argument is a version of the Aly-Sturrock conjecture \cite{Aly1991,Sturrock1991} which states that any equilibrium 
field will have lower energy than the corresponding open field.  Open fields are all asymptotically monopolar therefore cases with cartesian symmetry, such as ours, have infinite energy.  The field from \eq\ (\ref{eq:Fpo}) is not completely open since it contains closed flux in the side lobe domains 3 and 4.  This does not, however, prevent the energy from being unbounded and the state from being unachievable.

For an untethered field, including a domain 6 and without domain 2, we take $h\to\infty$ in the field expressed by \eq\ (\ref{eq:Fbc}).   In this case $u\to\infty$ with $h$, and all other quantities remain finite,
\be
  \hat{F}^{(pc)} ~=~ i\tilde{Q}{\sqrt{w^2+p^2}\sqrt{w^2+\tilde{q}^2}\sqrt{w^2+t^2}
  \over(w^2-x_1^2)(w^2-x_2^2)} ~~,
  	\label{eq:Fqp}
\ee
where $Du/h^2\to\tilde{Q}$.  The branch cut from $w=it$ extends vertically upward to infinity forming a sheet separating the upward from downward portions of domain 5.  

This {\em partially closed} field  is also asymptotically monopolar, and therefore has infinite energy as well.  This means that untethered equilibria are also stable, and that the transition from being tethered to untethered does not destabilize the system.  This is evident for the case of state $H$, which lies within region S of \fig\ \ref{fig:rx_phase}.  It also applies to cases where bifurcation occurs through a coalescence of $\tau$ and $-\tau^*$ within the the lower current sheet.  As this occurs the point moves off the right side of $(\theta,\Delta)$ space, but merely crosses the diagonal boundary in $(\psi_1,\psi_5)$ space.  Were equilibrium H to be driven into fold F2, say by breakout reconnection, it would drop into a stable equilibrium with a broken current sheet.

The ultimate eruption of the flux rope, in the sense of taking $h\to\infty$, therefore requires reconnection.  The existence of stable equilibria of the asymptotically octopolar form (see \fig\ \ref{fig:as_oct}) shows that the flux rope will not erupt unless 
$\psi_2\to0$, meaning the field must become untethered first.  Once untethered, only reconnection at the upper sheet, \ie\ breakout reconnection, will reduce $\psi_5\to0$ to permit eruption.  For the broken current sheet configuration, in which this occurs, breakout reconnection can occur in the section of the current sheet {\em below} the flux rope.  This will transfer flux out of domain 5, and into domain 6.  It will also add opposing flux to the closed flux surrounding the current filament.  By the time $\psi_5\to0$, and $\psi_6\to\psi_1-\pi(Q_2-Q_1)$, the closed azimuthal flux surrounding the current filament will have been reduced by this amount to $\psi_c\to A_h+\pi(Q_2-Q_1)$.  In cases where $A_h<-\pi(Q_2-Q_1)$ the flux rope will disappear ($I_0\to0$) before it manages to erupt; it vanishes when $\psi_5=\pi(Q_1-Q_2)-A_h>0$.  When this occurs the current sheet separating domains 5 and 6 vanishes too, and the domains become indistinguishable -- we can refer to it as domain 6.  In the other case, $A_h>-\pi(Q_2-Q_1)$ which applies to the case in \fig\ \ref{fig:psi15}, the flux rope remains in tact so $\psi_5\to0$ for the flux rope to erupt.

\begin{figure}[htp]
\centerline{\parbox[b]{2.3in}{\psfig{file=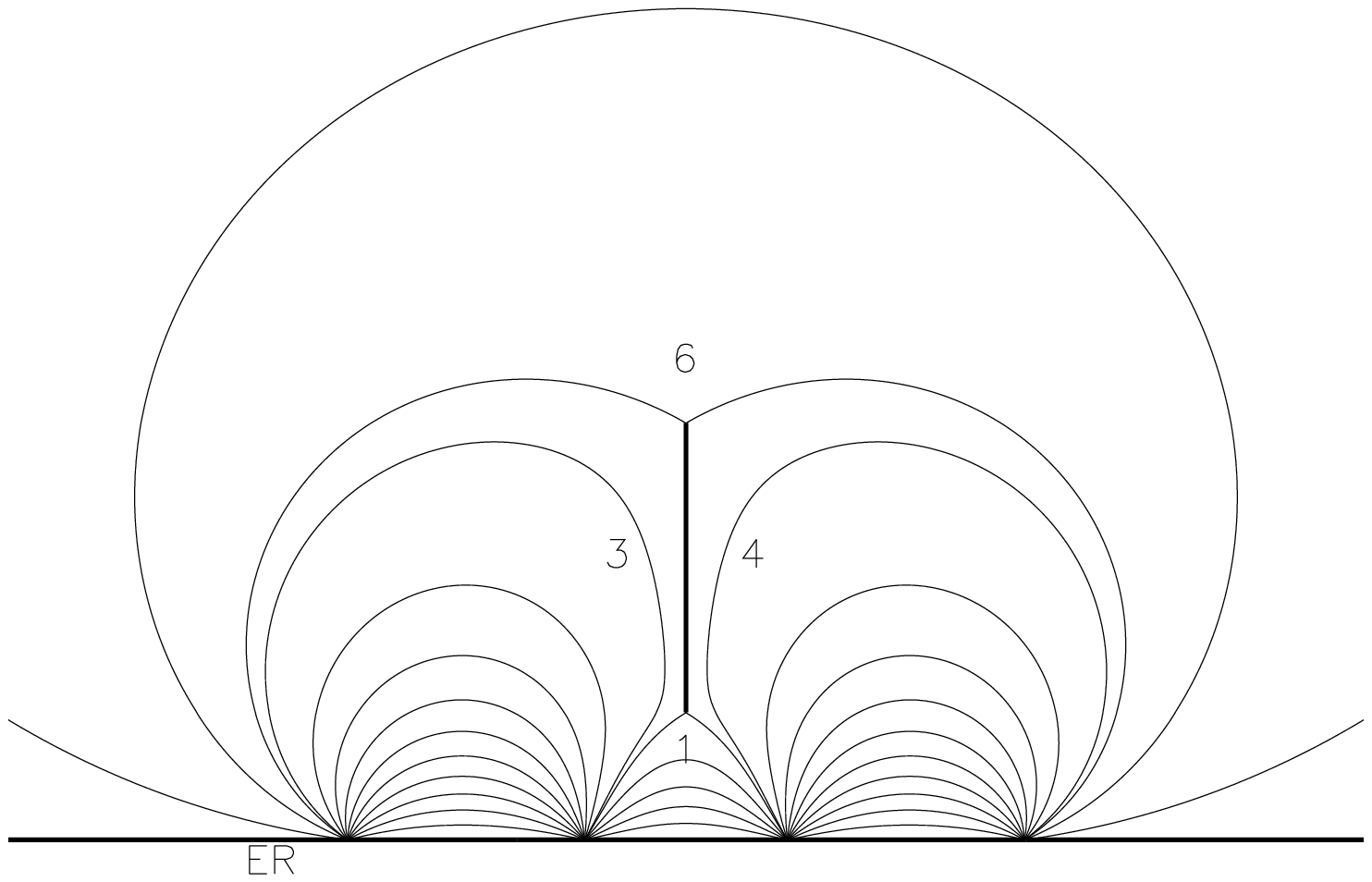,width=2.3in}\break
~~~~~~~~\psfig{file=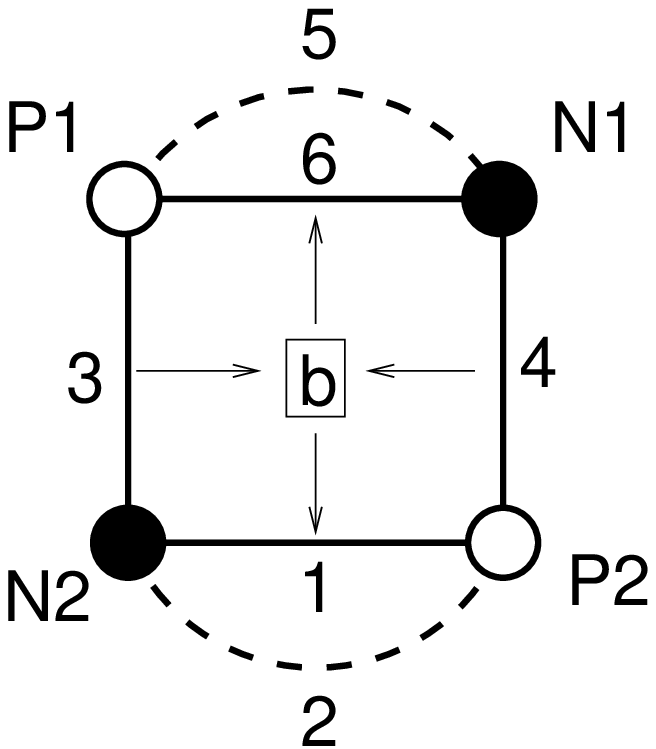,width=1.1in}}%
\psfig{file=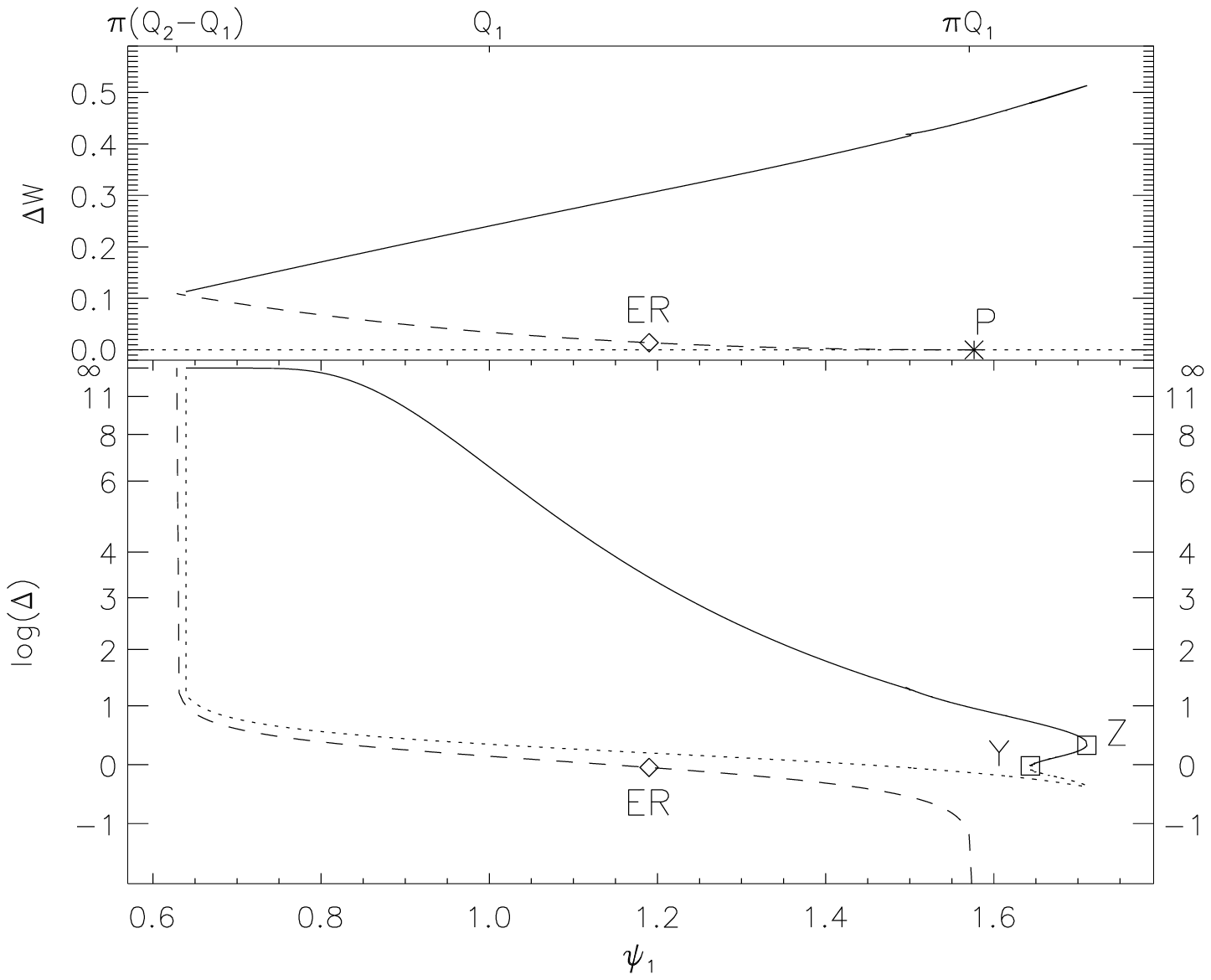,width=3.5in}}
\caption{Comparison of the marginally tethered field and the fully-erupted field.  The left panel shows an example (ER) of a fully erupted field.  It contains a single current sheet and domains 1, 3, 4, and 6 (labeled).  The graphs on the right compare the fully-erupted field (dashed) to the tethered field at the limit $\theta=\pi$ (solid).  The bottom panel compares the length $\Delta$ of their current sheets on a scale distorted to include $\Delta=\infty$.  Points labeled Y and Z correspond to points along the $\theta=\pi$ axis of \fig\ \ref{fig:map}.  The dotted curve is the length of the lower sheet, $\tilde{q}-p$, when the upper branch points have converged $\tau=-\tau^*=it=i\tilde{q}$.  The upper panel shows the free energy in both situations.  Point P is the potential field, $\hat{F}_0(w)$, with a single X-point and no current sheet.}
	\label{fig:po}
\end{figure}

Full eruption thus lies along the bottom axis ($\psi_5=0$) of \fig\ \ref{fig:rx_phase}, 
and must be reached by crossing the marginally tethered state ($\theta=\pi$) which is the edge of the shaded region.
This is generally approached at very large $\Delta$ --- far up the right axis of \fig\ \ref{fig:psi15}.  Figure \ref{fig:po} shows the length of the current sheet in this marginal case as a function of $\psi_1$.  Those few cases near equilibrium Z, with $\psi_1\approx 1.7Q_1$ (\fig\ \ref{fig:gap} is an example), have modest sheet 
lengths: $\Delta \approx x_1$.  For almost any smaller values of $\psi_1$ the case of marginal tethering is approached with a current sheet hundreds, thousand or even millions of times that long.  For $\psi_1\approx Q_1$ the length is 
$\Delta\approx 10^{7}\,x_1$.  An asymptotic analysis, presented in an appendix, shows that
\be
  \Delta ~\sim~ \exp\left[ \xi\, {(\psi_1+A_h)\over(\psi_1-\psi_{1,{\rm min}})}\,\right] ~~,
\ee
where $\psi_{1,{\rm min}}=\pi(Q_2-Q_1)$ is the minimum permissible value and $\xi$ is a constant.  This diverges extremely strongly as $\psi_1\to\psi_{1,{\rm min}}$, and it drives the flux rope along with it (the appendix also shows 
$h\approx \sqrt{3}\Delta$).  As large as this may be, it is nevertheless finite so the system has not, strictly speaking, undergone an eruption.  The asymptotic height is, however, so large that there is no hope that the evolution would remain quasi-static as $h$ was driven to values more than a million-fold larger than the underlying dipole.

Once the system has crossed into the untethered regime, the remaining flux overlying the filament can be reconnected.  This can be achieved only through breakout reconnection at the upper current sheet (although now its end is below the flux rope, as in \fig\ \ref{fig:gap}), transferring flux from domain 5 into domain 6.   When this reconnection runs to completion 
$\psi_5=0$ and the flux rope will have fully erupted: $h\to\infty$.  The resulting field is a classic quadrupole with no flux rope
\be
  \hat{F}^{(er)}(w) ~=~ i\tilde{D}{\sqrt{w^2+p^2}\sqrt{w^2+\tilde{q}^2}\over (w^2-x_1^2)(w^2-x_2^2)} ~~,
\ee
shown in \fig\ \ref{fig:po} (the super-script refers to ``erupted''.)  This is the limit of \eq\ (\ref{eq:Fbc}) when $h$, $t$, and $u$ all go to infinity and $Dtu/h^2\to\tilde{D}$ is the new dipole moment.  The field is asymptotically dipolar ($\hat{F}\sim i\tilde{D}/w^2$) and thus has bounded free energy (the dashed curve on the top right of \fig\ \ref{fig:po}).  As $\psi_1\to\pi(Q_2-Q_1)$ in the marginally tethered case $u\sim h\to\infty$, so $\hat{F}^{(bc)}\to\hat{F}^{(er)}$, and the energies of the two states match.  For every value of 
$\psi_1>\pi(Q_2-Q_1)$, the breakout reconnection will release significant energy as it drops the system into the erupted state.

Tether-cutting reconnection, on the other hand, will increase $\psi_1$ by transferring flux from the side lobes 3 and 4.  This will reduce the free energy (dashed curve in the upper right panel of \fig\ \ref{fig:po}), as the field approaches the potential state P ($p=\tilde{q}$, and the current sheet becomes an X-point).  It is clear from the free energy plot that except in the vicinity of $\psi_1=\pi(Q_2-Q_1)$, the tether-cutting reconnection releases far less energy than the breakout reconnection; \ie\ the difference between solid and dashed curves is far greater than the distance between the dashed and $\Delta W=0$.

\subsection{Catastrophic Instability Triggered by Ideal Evolution}

The foregoing discussion focussed on how reconnection at either of the current sheets can lead, under some circumstances, to a loss of equilibrium.  The alternative scenario is ideal evolution in which no flux is transferred across the current sheet.  One example, photospheric shearing motion adding axial flux to the flux rope, is not possible within our strictly two-dimensional model.  A similar effect can, however, be achieved by adding closed azimuthal flux to the flux rope.  This is accomplished by increasing the value of $A_h$, while holding fixed $\psi_1$.  Like the shearing motion, this adds magnetic pressure to the flux rope which pushes against the overlying field.

\begin{figure}[htp]
\psfig{file=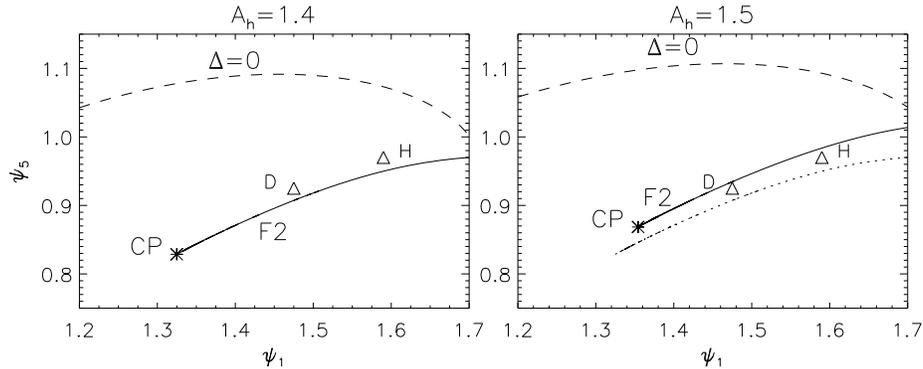,width=5.1in}
\caption{Detail from the upper right corner of \fig\ \ref{fig:rx_phase}, showing the fold $F2$ (solid), the critical point CP (asterisk), and the upper portion of the $\Delta=0$ boundary (dashed).  The left panel shows the configuration for the same parameters as \fig\ \ref{fig:rx_phase}: $A_h=1.4Q_1$.  The right panel shows the situation when $A_h=1.5Q_1$.  The critical point and fold have moved upward and rightward; previous location is indicated with a dotted curve.  Locations of equilibria D and H, indicated by triangles, remain unchanged.  The motion of the fold causes both equilibria to undergo loss of equilibrium.}
	\label{fig:ideal_loe}
\end{figure}

Figure \ref{fig:ideal_loe} shows the effect of this kind of driving.  Equilibrium points, such as H and D (shown previously in \figs\ \ref{fig:disc} and \ref{fig:loe} respectively) remain fixed in $(\psi_1,\psi_5)$ space, while the structure of the equilibrium space evolves around them.  In this case the critical point, CP, and fold F2, move up and right toward larger values of 
$\psi_1$ and $\psi_5$.  This motion achieves the same end as reconnection would: it creates a loss of equilibrium in both H and D.  The fold moves past them depriving them of their second stable equilibrium.  The result is that they must expand rapidly upward, as illustrated in \fig\ \ref{fig:loe}.  The only difference is that in this case the fold moves across the stationary equilibrium while reconnection moves the equilibrium downward or rightward across the stationary fold.

A similar effect could be achieved by the subduction of the inner dipole.  This would be achieved by decreasing $Q_2$, and decreasing $\psi_1$ at the same rate.  The closed flux in the rope, $\psi_c=A_h-\psi_1$, would remain fixed so $A_h$ would {\em increase} during subduction.



\section{Discussion}

We have presented an analytic model of equilibrium current filaments anchored to quadrupolar flux distributions.  These are flux constrained equilibria \cite{Longcope2001b} with two current sheets whose size and net current depends on the distribution of domain fluxes.  We performed a thorough analysis of the equilibrium space for one particular set of equilibria: $x_2/x_1=0.3$, $Q_2/Q_1=1.2$, with an idealized current filament of radius $R=0.005x_1$.  We found a two-dimensional space of equilibria matching these conditions.  When these are characterized in terms of domain fluxes there are regions with multiple equilibria consistent with the same fluxes.  These regions are bounded by folds or cusps, at which loss of equilibrium can occur through quasi-static evolution driven by reconnection at either current sheet, or by ideal evolution.

We do not present detailed analyses of other configurations in this work, although several we have examined are qualitatively similar to the case presented.  It was shown in Section 2 that all equilibria with two distinct coronal X-points are unstable under vertical displacement of the line current.  In the case we presented, as in all other cases, an upward displacement will result in a new stable equilibrium with currents sheets above and beneath the line current, of the kind we have considered here.  Thus at least the lower left corner of equilibrium space will be qualitatively similar for all other configurations.

In order to make use of analytic complex functions, our model assumed Cartesian rather than cylindrical symmetry.  Since general open-field equilibria are asymptotically monopolar they have infinite energy in Cartesian symmetry ($|\bvec|\sim r^{-1}$) but finite energy in cylindrical symmetry ($|\bvec|\sim r^{-2}$).  Models with cylindrical symmetry can therefore produce genuine eruption, with the current filament going to infinity and dragging open some flux.   Due to energetic constraints, flux cannot be opened in Cartesian models so they produce only a finite displacement to a new stable equilibrium.  Our analysis found the LOE generally occurs at current filament heights comparable to the extent of the photospheric flux distribution (\ie\ $x_1\sim 10^{10}$ cm).  At these modest heights the curvature effects of the cylindrical geometry would have been small if they were present, so we expect the LOE itself to be similar in both cases.  It is in the second equilibrium, which we have shown to be thousands or millions of times higher than the unstable one, where the absence of curvature, and therefore the limitations of Cartesian symmetry, become significant.  We therefore believe that the LOE we find in our Cartesian model would occur in a cylindrical model, but as a genuine eruption.

Our model demonstrates that LOE can be triggered either by reconnection at the upper current sheet (breakout reconnection) or by reconnection at the lower current sheet (internal tether cutting).  This suggests that it is not possible to dismiss {\em a priori} either scenario as an eruption mechanism.  We do find, however, that some flux distributions can be destabilized by one mechanism and not the other.  In order to apply this insight to simulations with cylindrical symmetry, it will be necessary to recast our analysis in that form.  Thus we cannot, at present, apply our model directly to simulations such as Karpen, Antiochos, and Devore (2012).\nocite{Karpen2012}  Their finding that the simulated eruption is triggered by breakout reconnection, and not by tether cutting, is consistent with a LOE from our analysis.

The equilibrium space we analyzed had a single curve (F2) across which an LOE would occur through quasi-static evolution.  It is hard to envision a scenario in which the system crosses this curve twice in succession to produce the kind of double catastrophe observed by Zhang, Hu, and Wang (2005).\nocite{Zhang2005}  Their model, however, had cylindrical symmetry and included an azimuthal  field component.  It is therefore possible that this equilibrium space is more complex than the two-dimensional Cartesian system we analyzed.

Our model of magnetic reconnection was kinematic and quasi-static: a  small reconnection electric field transferred flux across the current sheet slowly enough that the system could remain in mechanical equilibrium.  This reveals how much magnetic energy could be released by reconnection, but does not consider the fate of the released energy.  We found that over much of the equilibrium space a reconnection electric field in a current sheet resulted in an increase in the size and strength of the current sheet.  Some dynamical reconnection mechanisms, such a the tearing instability \cite{Furth1963} or ion-acoustic turbulence \cite{Somov1985}, become more effective as the current sheet becomes longer or stronger. Our two-current-sheet model would provide a positive feedback to such a mechanism through its large-scale dynamics, causing the small-scale mechanism to run away.  

The reasoning above suggests the possibility of a resistive eruption mechanism, although it lies outside the scope of the present investigation.  Instead we have demonstrated ideal instabilities which occur through the loss of equilibrium.  Although the instability is ideal, driven by an imbalance in Lorentz forces, it can be triggered either by ideal motions or by reconnective flux transfer.  It will occur only at the fold, F2, while a resistive instability, caused by run-away microscopic reconnection, could occur at any point in equilibrium space where the global field provided positive feedback.

\bigskip

This work was supported in part by a grant from the NSF/DOE Plasma Sciences partnership.

\appendix

\section{Locating the Upper Current Sheet}

In the neighborhood of the right branch point, $w=\tau$, the complex function in 
\eq\ (\ref{eq:F}) takes the form
\be
  \hat{F}(w) ~\approx~ \zeta_{\tau}\sqrt{w-\tau} ~~,
\ee
where $\zeta_{\tau}$ is a complex number found from the limit.  The point $w=\tau+\epsilon e^{i\phi}$ is displaced from the branch point a small distance, $\epsilon$, in direction $\phi$.  The direction of the magnetic field vector at this point is given by the complex phase of
\be
  B_x+iB_y ~=~ i\hat{F}^* ~=~ i\zeta_{\tau}^*\,\sqrt{\epsilon}\,e^{-i\phi/2} ~~.
\ee
For the direction of the field to be parallel or anti-parallel to the displacement direction, 
$\phi$, requires
\be
  {\rm Im}\{\ln(B_x+iB_y)\} ~=~ \hbox{${\pi\over2}$}-{\rm Im}\{\ln\zeta_{\tau}\}
  - \half \phi ~=~ \phi + n\pi ~~,
\ee
for some integer $n$.  The branch cut should attach to the branch point along a direction satisfying this requirement, thus it leaves along the direction
\be
  \phi ~=~\hbox{${\pi\over3}$}-\hbox{${2\over3}$}{\rm Im}\{\ln\zeta_{\tau}\}
  - \hbox{${2\over3}$}\pi n ~~.
\ee
One of the choices $n=0,1$, or 2 will be the branch cut containing the current sheet, while the other two directions are separatrices connecting to the branch point.  It is clear that these three form an equiangular $Y$-point.

For the case of a very small current sheet $\tau\approx iy_a$, where $y_a>h,p,q$, is the X-point of the potential field.  In this case
\[
  \zeta_{\tau} ~=~iD{\sqrt{(\tau^2+p^2)(\tau^2+q^2)}\sqrt{2\tau(\tau^2-\tau^{*2})}\over
  (\tau^2-x_1^2)(\tau^2-x_2^2)(\tau^2+h^2)} ~\approx~
  D{\sqrt{(y_a^2-p^2)(y_a^2-q^2)}\sqrt{8 y_a^2 r}\over
  (y_a^2+x_1^2)(y_a^2+x_2^2)(y_a^2-h^2)} ~~,
\]
where $r>0$ is the small real part of $\tau$.  When $D<0$, as it will be when 
the far field is rightward, we find ${\rm Im}\{\ln\zeta_{\tau}\}=\pi$ and the angle 
of the current sheet is $\phi\approx-\pi$ when we choose $n=1$.  This is the 
horizontal current sheet while the $n=0$ and $n=2$ branches are the separatrices from it.

For the correct choice of branch cut direction a field line can be integrated from near the branch point, 
$w=\tau+\epsilon e^{i\phi}$, to the mid-plane ($x=0$) where it will be 
horizontal ($B_y=0$).  By symmetry this field line will continue across the mid-plane and connect to the other branch point at $w=-\tau^*$.  The complete curve is the branch cut, $\bc_a$, across which 
$\hat{F}(w)$ will change sign.  If we begin with a traditional interpretation of the factor 
$\sqrt{(w-\tau)(w+\tau^*)}$, commonly in numerical implementations, the branch cut 
will be a straight horizontal line, ${\cal H}_a$, between $\tau$ and $-\tau^*$.  To construct the correct version from this we change the sign of $\hat{F}(w)$ whenever $w$ lies inside the closed curve  ${\cal H}_a\cup\bc_a$; this has the effect of moving the 
branch cut ${\cal H}_a\to\bc_a$.
The current in this upper current sheet is found by integrating along the branch cut
\be
  I_a ~=~ - {1\over 2\pi}\int_{\bc_a}|\hat{F}(w)|\, d\ell ~~,
\ee
where $d\ell$ is the length element along the branch cut.

\section{Asymptotics of the Marginally Tethered State}

In the limit approaching eruption both $h$ and $q$ will be very large compared to $p$, $\tau$, $x_1$, and $x_2$.  In the vicinity of the flux rope $w$ will also be large and \eq\ (\ref{eq:F}) approaches
\be
  \hat{F}(w) ~\approx~ {iD\over w}\,{\sqrt{w^2+q^2}\over w^2+h^2} ~~.
\ee
Force balance, \eq\ (\ref{eq:B0}),
\be
  \left.{d\over dw}\left[ (w-ih)\hat{F}\right]\, \right\vert_{w=ih} ~=~ {D\sqrt{h^2-q^2}\over h^2}\left[ {3\over 2h}
  - {h\over h^2-q^2} \right] ~=~ 0 ~~,
\ee
requires that $h=\sqrt{3}q$.  The closed magnetic flux passing between $y=q$ and $y=h-R$ is
\be
  \psi_c ~=~ \int\limits_q^{h-R}{D\sqrt{y^2-q^2}\over y(h^2-y^2)}\,dy ~\approx~
  {D\over \sqrt{6}h}\ln(c_0 h/R ) ~~,
  	\label{eq:psic_asymp}
\ee
where $c_0=(4/3)\exp[-\sqrt{2\,}\hbox{atan}(\sqrt{2})]$.

In the region far below the flux rope, $|w|\ll q$, the marginally tethered case 
$\tau=-i\tau^*=it$, has a field given by
\be
  \hat{F} ~\approx~ {iDq\over h^2}{\sqrt{w^2+p^2}(w^2+t^2)\over (w^2-x_1^2)(w^2-x_2^2)} ~~.
\ee
Equations (\ref{eq:Q1}) and (\ref{eq:Q2}) become, in this limit
\begin{eqnarray}
  {Dq\sqrt{p^2+x_1^2}(x_1^2+t^2)\over 2h^2x_1(x_1^2-x_2^2)} &=& Q_1 ~~,\\
  {Dq\sqrt{p^2+x_2^2}(x_2^2+t^2)\over 2h^2x_2(x_1^2-x_2^2)} &=& Q_2  ~~.
\end{eqnarray}
These can be combined to find
\be
  p ~=~ x_1x_2\,\left[ {Q_2^2(t^2+x_1^2)^2-Q_1^2(t^2+x_2^2)^2\over
  Q_1^2x_1^2(t^2+x_2^2)^2-Q_2^2x_2^2(t^2+x_1^2)^2} \right]^{1/2} ~~,
  	\label{eq:p_asym}
\ee
leading to
\be
  {Dq\over h^2} ~=~ 2Q_1x_1{x_1^2-x_2^2\over \sqrt{p^2+x_1^2}(t^2+x_1^2)} ~~.
\ee
Finally, the flux in domain 1 is determined by the integral
\be
  \psi_1 ~=~ {Dq\over h^2}\int\limits_0^{p}{\sqrt{p^2-y^2}(t^2-y^2)\over
  (y^2+x_1^2)(y^2+x_2^2)} \, dy ~~.
    	\label{eq:psi1_asym}
\ee

If the current sheet break is far from the surface, $t\gg x_1$, but still far below the flux rope, $t\ll q$, 
\eq\ (\ref{eq:p_asym}) becomes
\begin{eqnarray}
  p &\approx& x_1x_2\,\sqrt{Q_2^2-Q_1^2\over
  Q_1^2x_1^2-Q_2^2x_2^2} 
 +{c_1\over t^2} ~+~\cdots ~=~  p_{\infty} ~+~ {c_1\over t^2} ~+~\cdots ~~,
\end{eqnarray}
where $c_1$ is a constant.  In the same limit
\be
  {Dq\over h^2} ~\approx~ {2Q_1(x_1^2-x_2^2)\over x_1\sqrt{p_{\infty}^2+x_1^2}}\,{1\over t^2} +\cdots
  ~=~ {c_2\over t^2} + \cdots ~~.
  	\label{eq:Dq_asymp2}
\ee
and the domain flux is 
\be
  \psi_1 ~\approx~ {Dq\over h^2}t^2\int\limits_0^{p_{\infty}}{\sqrt{p_{\infty}^2-y^2}\over
  (y^2+x_1^2)(y^2+x_2^2)} \, dy ~+~ {c_3\over t^2} + \cdots ~=~ 
  \psi_{1,{\rm min}}~+~ {c_3\over t^2} + \cdots ~~.
    	\label{eq:psi1_asymp2}
\ee
The first integral can be evaluated, but it is also possible to see that $\psi_{1,{\rm min}}=\pi(Q_2-Q_1)$ since with 
$h\to\infty$ all flux from N1 must connect to P2, and the remaining flux from P2 forms domain 1 by connecting to N1.

Equations (\ref{eq:Dq_asymp2}) and (\ref{eq:psi1_asymp2}) can be combined to yield
\be
  {Dq\over h^2} ~=~ {c_2\over c_3}\,(\psi_1-\psi_{1,{\rm min}}) ~~.
\ee
Using this in \eq\ (\ref{eq:psic_asymp})
\be
  \psi_c ~=~ {c_2\over \sqrt{2}c_3}\,(\psi_1-\psi_{1,{\rm min}})\,\ln(c_0 h/R) ~~.
\ee
The closed flux contributes to the constant: $A_h=\psi_c-\psi_1$.  This can be used to obtain the flux rope height
\be
  h ~=~ {R\over c_0}\,\exp\left[ {\sqrt{2}c_3(\psi_1+A_h)\over c_2 (\psi_1-\psi_{1,{\rm min}})}\right] ~~.
\ee
The height diverges extremely rapidly as $\psi_1\to\psi_{1,{\rm min}}$.


\end{article}
\end{document}